\documentclass[conference]{acmsiggraph}

\usepackage{paralist}
\usepackage{xcolor}
\usepackage{multirow}
\usepackage{tabularx}
\usepackage{graphicx}
\usepackage{booktabs}
\usepackage{multirow}

\title{\emph{SculptStat}: Statistical Analysis of Digital Sculpting Workflows}

\author{Christian Santoni\thanks{e-mail: \{surname\}@di.uniroma1.it} \hspace{0.1in} Claudio Calabrese\footnotemark[1] \hspace{0.1in} Francesco Di Renzo\footnotemark[1] \hspace{0.1in} Fabio Pellacini\footnotemark[1] \\Sapienza University of Rome}

\pdfauthor{Christian Santoni}

\newcommand{\parasection} [1] {\textsf{\textbf{#1}.}}

\newcommand{\oldversion } [1] { }
\newcommand{\ignorethis } [1] { }

\newcommand{\tblnum     } [1] {\ref{#1}}
\newcommand{\fignum     } [1] {\ref{#1}}

\newcommand{\tbl        } [1] {Tab.~\tblnum{#1}}

\newcommand{\fig        } [1] {Fig.~\fignum{#1}}

\newcommand{\Reals      }     {{\textrm{I\kern-0.18em R}}}

\newcommand{\change     } [1] {\mbox{{\footnotesize $\Delta$} \kern-3pt}#1}

\newcommand{\meshwidth}{6in}
\newcommand{\tblscale}{0.8}
\newcommand{\diagramwidth}{3.1in}

\begin{document}

\teaser{
    \begin{center}
    \includegraphics[width=\meshwidth]{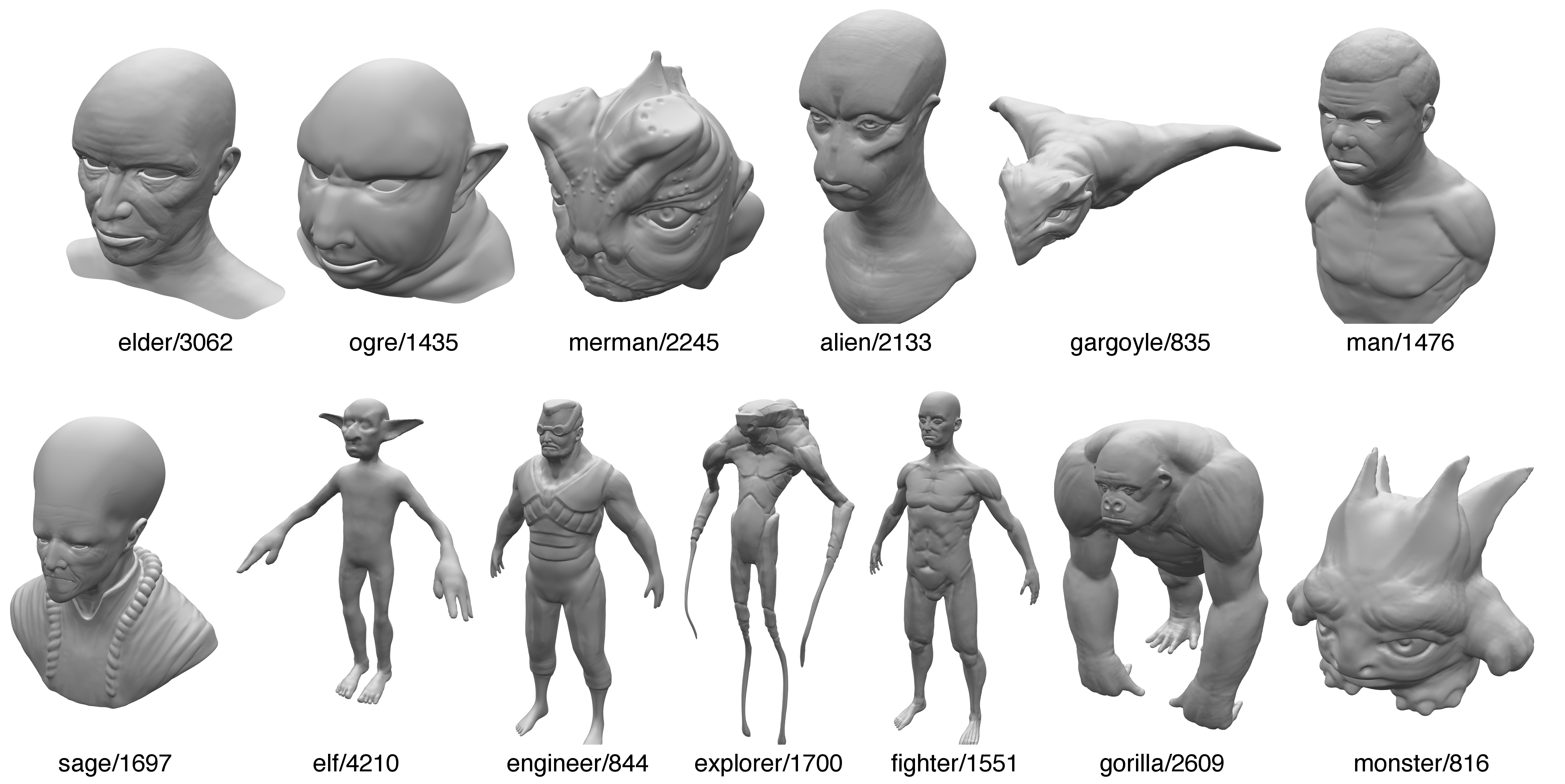}
    \end{center}
    \caption{Models analyzed in the paper. Expert artists create the models with the specified number of strokes.}
    \label{fig:teaser}
}

\maketitle


\begin{abstract}
Targeted user studies are often employed to measure how well artists can perform specific tasks. But these studies cannot properly describe editing workflows as wholes, since they guide the artists both by choosing the tasks and by using simplified interfaces. In this paper, we investigate digital sculpting workflows used to produce detailed models. In our experiment design, artists can choose freely what and how to model. We recover whole-workflow trends with sophisticated statistical analyzes and validate these trends with goodness-of-fits measures. We record brush strokes and mesh snapshots by instrumenting a sculpting program and analyze the distribution of these properties and their spatial and temporal characteristics. We hired expert artists that can produce relatively sophisticated models in short time, since their workflows are representative of best practices. We analyze 13 meshes corresponding to roughly 25 thousand strokes in total. We found that artists work mainly with short strokes, with average stroke length dependent on model features rather than the artist itself. Temporally, artists do not work coarse-to-fine but rather in bursts. Spatially, artists focus on some selected regions by dedicating different amounts of edits and by applying different techniques. Spatio-temporally, artists return to work on the same area multiple times without any apparent periodicity. We release the entire dataset and all code used for the analyzes as reference for the community.
\end{abstract}

\section{Introduction}

\parasection{Content Creation}
The human effort necessary to create 3D digital content is a major limiting factor for graphics applications, even today when modeling software and acquisition methods have improved significantly. A deep understanding of how artists create digital content is crucial in improving modeling workflows. User studies are employed to measure the performance of artists when performing creation tasks. The majority of these studies though are designed to validate particular algorithms rather than to characterize workflows as a whole.

\parasection{\emph{SculptStat}}
In this paper, we \emph{statistically characterize digital sculpting workflows by analyzing artists' behavior as they freely sculpt organic models}. In sculpting, artists alter the model's geometry by adopting an approach that is similar to clay sculpting, modifying parts of the model (i.e., vertices) using a set of different tools, whose effects span from minor surface modifications (like smoothing) to moving or extruding whole parts of the model. We chose to investigate sculpting since it is often used to define the shape of organic objects, such as characters, and since we are not aware of comprehensive user studies on this subject. We focus on detailed models, rather than simple edits, where overall workflow characteristics become apparent. Our experimental methodology is in stark contrast with most published literature in graphics, e.g. \cite{lighting,materials,lightfields}.

\parasection{Experiment}
We avoid using targeted experiments that would be limited to only a few aspects of sculpting and therefore provide no insight in the workflow as a whole. Instead, we let artists choose freely what to model and how. We only ensure that the chosen models span different types (head, bust, body), and that artists work both from scratch or base meshes. We analyze the data with sophisticated statistical methods that let us interpret this heterogeneous data. In contrast, most prior work guides artists with short tasks that have simple goals, requiring a relatively unsophisticated analysis. But workflows cannot be characterized with this data alone. 

To keep datasets and analysis to a manageable size, we characterize workflows across model types, leaving inter- and intra-artist investigations to future work. Furthermore, we do not know how to quantitatively compare subjects without guidance, since results in the literature show that open studies are not a reliable source for inter-subject comparisons (e.g. \cite{lighting,materials}).

To produce high quality models, we hired two experts, one of which is also an instructor, that are renowned in their communities. These artists can produce high quality models in relatively short workflows. We chose these experts since their workflows are representative of best practices, and since they teach such practices to others. We avoid investigating novices since they would not be able to produce the models in this paper without guidance in the experiment and interface.

We adopt Blender \cite{blender} as our sculpting interface since it is freely available, has a strong community, and is mature. The hired artists have significant experience with it. Blender's sculpting toolset is similar enough to other sculpting packages \cite{zbrush,mudbox} that we can consider the studied workflows representative. In contrast, prior work often uses simplified interfaces that are both easier to analyze and provide guidance to users, but subjects have no significant experience in this case, hindering their creative process. 

\parasection{Analysis}
We record brush properties, 3D and 2D locations and tablet pressures of strokes, and mesh snapshots after each stroke. On this data, we perform a variety of quantitative and qualitative analyses to characterize the artists' workflows. We analyze overall trends by determining the statical distributions of stroke properties (e.g., length, size, and angles) and mesh differences between subsequent strokes. We investigate temporal behaviors by modeling workflows as time series, performing temporal filtering, fitting to autoregressive moving average models, and estimating hidden Markov models. We characterize spatial behaviors by sampling strokes as point clouds and analyze their density distributions as well as the correlation of brush properties with spatial positions. Finally, we consider spatio-temporal patterns by performing spatio-temporal clustering with ST-DBSCAN \cite{stdbscan}, a density-based clustering algorithm. We validate all these analyses with tests for statistical significance and quality of fit.
 
\parasection{Results and Contributions}
We analyzed 13 models, shown in \fig{fig:teaser}. These models took between 816 and 4210 strokes to be created, showing that experts take relatively little time to create complex models with today's interfaces, and giving us a very large dataset to analyze. The main findings of our work are:
\begin{compactenum}
    \item stroke length, brush size, and mesh differences are described well with inverse gaussian distributions, showing a preference for shorter and smaller strokes, with a long tail of longer ones; average stroke length and brush size are related to model features, not artist's style;
    \item temporally, artists do not work in a coarse-to-fine fashion, but in bursts of activity on different parts of the mesh;
    \item spatially, artists spend most of the time on some selected areas, but spatially distributing their edits in less than expected peaked distributions. Also, strokes' parameters are correlated to spatial locations (i.e.: similar values for such parameters are spatially clustered);
    \item artists return to the same model regions multiple times, in spatio-temporal clusters of activity.
\end{compactenum}
To aid in reproducibility and foster future research, we release all stroke and mesh data, the code used for all statistical analyses, and further documentation as supplemental (we did not include meshes for review since the compressed dataset takes 100 GB). In conclusion, the contributions of our work are:
\begin{compactenum}
    \item we establish a methodology, based on solid statistical foundations, to analyze complex 3D content creation workflows;
    \item we precisely characterize digital sculpting workflows, deriving many previously unknown trends;
    \item throughout the paper, we provide interpretations of such analysis and make suggestions for future tools development;
    \item we provide a large dataset of sculpting data to the community, and all tools we used for the analysis.
\end{compactenum}


\section{State of the Art}

\parasection{Workflow analysis}
The idea of analyzing workflows is certainly not new and the literature provides many examples supporting such interest, especially in fields close to computer graphics and human-computer interaction. For example, \cite{ilovesketch,drawlines} investigates how people draw lines, \cite{3dcurve} discusses how to sketch 3d surfaces, \cite{global,lighting,materials,lightfields} quantitatively evaluate the efficiency of several interface paradigms for lighting, material editing and light fields. In general though, all these works focus on studies that answer specific design questions. We have a diametrically opposed goal, in that we want to characterize whole workflows from unguided experiments.

The works that are mostly related to ours are the ones that use design workflows as main subject of research, including \cite{delta} for image editing and \cite{slides} for slideshows. While the basic idea is similar, our interest is not focused on comparing directly two artists' workflows, but to have a better insight on the general trends and behaviours which artists make use of while creating their models. To the best of our knowledge, there is no published work focusing on the study of 3D editing workflows, while this analysis has been already conducted for 2D images, like in \cite{berger}. Other works, like \cite{3dauthoring} for sculpting and \cite{meshflow,vistrails} for polygonal meshes, provide visual summaries of workflows to aid artists in understanding each other's techniques, but do not produce any quantitative measure that could characterize the workflow itself.

\parasection{Editing sequences}
We use software instrumentation to record the full editing history. Other works have attempted to recover such histories from edit snapshots, i.e. \cite{inverse} for images and \cite{3dtimeline} for meshes. These inverse editing approaches work well for simple operations, but fail to capture editing semantic for larger changes. More importantly, they do not capture designers' actions. Recent works focus on version control frameworks to store, diff and merge editing sequences, e.g. \cite{meshgit,3ddiff} for meshes and \cite{nonlinear} for images. These approaches aim to be efficient and independent of editing operations, which is in contrast with our requirement to track all possible user actions.


\begin{figure*}
    \begin{center}
    \includegraphics[width=\meshwidth]{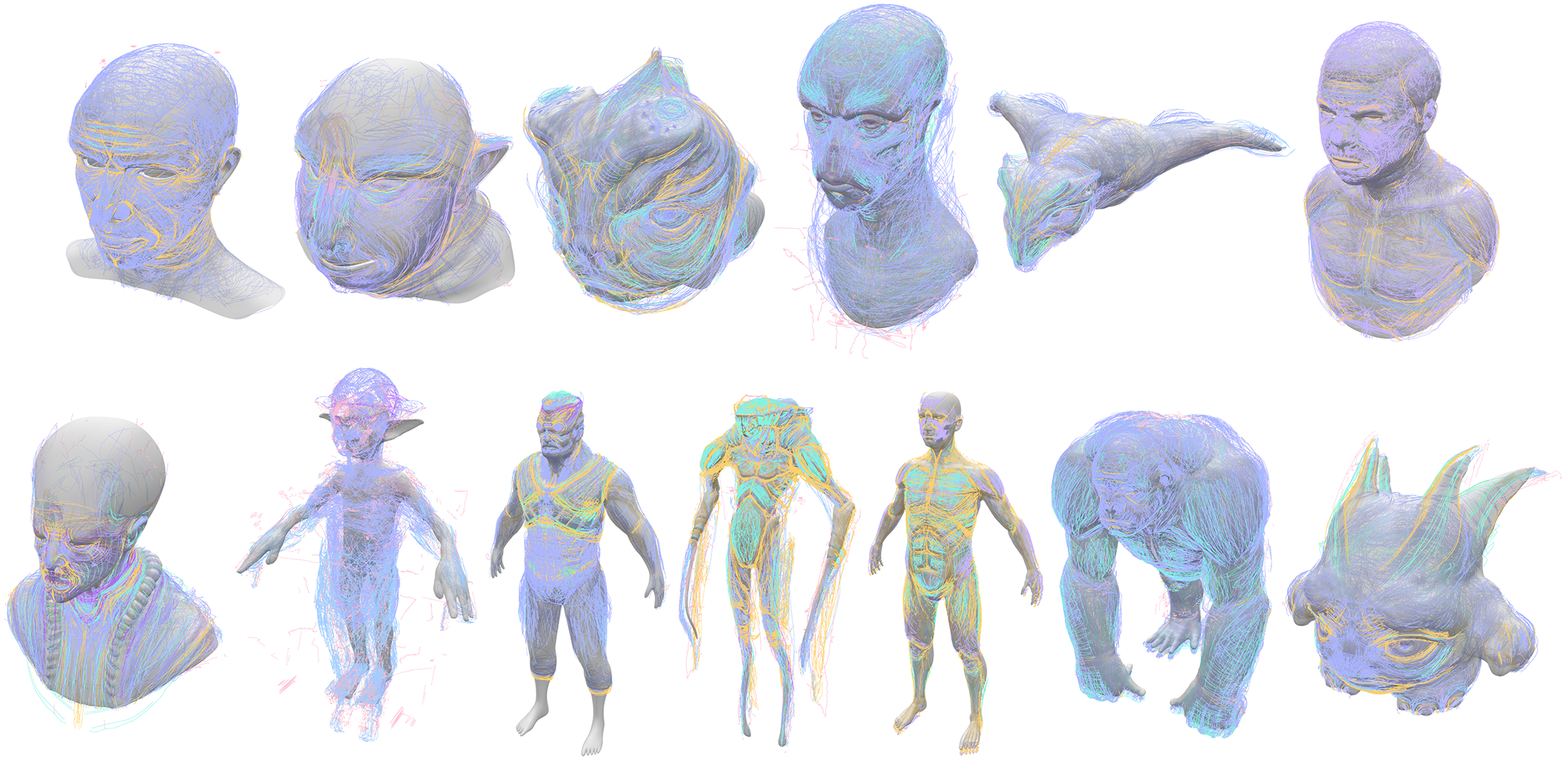}\\
    \includegraphics[width=\meshwidth]{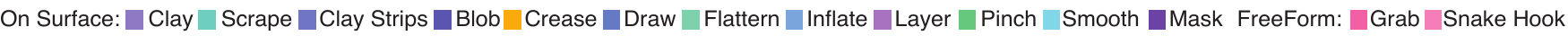}
    \end{center}
    \caption{Strokes used to create the models rendered over a semi-transparent mesh as reference. Color indicate brush type. We report brush names as displayed in the interface. The mesh is rendered slightly smaller to aid in visibility. Note that some strokes appear not on the surface since overall mesh shape was modified after the strokes were applied.}
    \label{fig:strokes}
\end{figure*}

\section{Dataset}

\parasection{Models}
We statistically characterize digital sculpting workflows when modeling organic characters. Within this model category, we let users freely choose what and how to model. \fig{fig:teaser} shows the models created for this paper. \tbl{tbl:models} summarizes statistics for the models. In general, we asked subjects to model organic characters of their choosing, spanning a variety of techniques using Blender. We considered heads, busts and full-bodies modeling, starting from scratch or using base meshes. Base meshes are shown in supplemental. Artists sculpted using subdivision modeling or dynamic topology \cite{freestyle}. Models took between 29 minutes and 4 hours to create.

\parasection{Subjects}
We hired two experts to produce the models in this paper. We choose experts since they can create sophisticated models in relatively short time, without guidance in the experiment and interface. To limit the experiment to a manageable size, one can choose to either pick many subjects, each of which creates a model, or focus on a few subjects that create many models. We choose the latter since we  wanted to establish a statistical methodology to characterize workflows from ``freely-captured'' data, validated across model types, rather than investigating inter- or intra-subject analysis. We chose subjects that are known in their community and are accomplished instructors. Their workflows are representative of best practices and what others are taught to do. Note that this practice is accepted in statistics.

\parasection{Captured Data}
We instrumented Blender to store all user actions and save mesh snapshots after each stroke. The instrumentation is fast enough to leave workflow unaffected. The models in this paper were created with between 816 and 4210 brush strokes and between 225 and 956 camera changes. This number of strokes is sufficient for detailed models. All models were created with mirrored strokes, and overall model size was left to the artists. We focus our analysis on the strokes since camera changes have been considered in \cite{3dauthoring}.

\parasection{Brushes and Strokes}
Blender uses a brush model that takes inspiration from works like \cite{sweeper,freestyle}. Brushes have many parameters that control their behavior, including size, falloff and type. Brush types describe what geometric deformations are applied to the model and the behavior when strokes overlap. \ignorethis{\fig{fig:strokes} show a few brush types.} We found that artists rarely change brush parameters, so we focus our analysis only on brush size and type. We represent strokes as the polylines sampled by the interface, with positions in model space and tablet pressures at polyline vertices. Brush sizes are in model space, rather the 2D screen pixels, since we found no significant trends in the latter. \fig{fig:strokes} shows the strokes needed to generate the models.

\parasection{Brush Types}
We consider two groups of brush types: on-surface and freeform. For on-surface brushes, the interface determines the 3D position of a stroke point by projecting the 2D mouse location on the mesh. For freeform strokes, the 3D position is derived by the 2D mouse location by moving the point on a plane parallel to the camera. On-surface brushes are typically used to push or pull vertices along the surface normals or to add or remove volume like clay sculpting. Freeform brushes are mostly used to extrude new parts of the model or to apply freeform deformation to large areas. 

\parasection{Mesh Distances}
In general, there is no simple relation between brush and stroke parameters and the corresponding mesh difference \cite{sweeper}. To measure the magnitude of the effect of a stroke, we compute the Hausdorff distance between the mesh before and after the stroke was applied, using \cite{metro}. We chose this metric since it correlates well (with an average Pearson's \textit{r = 0.99}, computed with a Fisher \textit{z} transformation of the individual \textit{r} of each model) with a simple estimate of stroke effect, computed as the product of stroke length, brush size and average pressure.

\parasection{Reproducibility}
To aid in further analysis, the \emph{supplemental material includes all captured data (strokes and meshes distances), the code used to perform the analysis, and detailed per-model diagrams}.

\begin{table}[tb!]
\begin{center}
\scalebox{\tblscale}{
\begin{tabular}{lcrrrr}
\toprule
\multicolumn{1}{c}{\textbf{Model}} & \multicolumn{1}{c}{\textbf{Artist}} &   \multicolumn{1}{c}{\textbf{Time}} & \multicolumn{1}{c}{\textbf{Strokes}} & \multicolumn{1}{c}{\textbf{Camera}} & \multicolumn{1}{c}{\textbf{Size}} \\
\midrule

alien & A    &  1h12m &     2133 &        660 &       4.8841 \\
elder & A    &  1h24m &     3062 &        519 &      18.1198 \\
ogre & A     &  0h35m &     1435 &        324 &      20.0907 \\
merman & B  &  0h58m &     2245 &        714 &       4.9031 \\
man & A      &  1h11m &     1476 &        462 &       4.3839 \\
monster & B  &  0h40m &      816 &        492 &       6.8777 \\
sage & A     &  1h04m &     1697 &        793 &       4.9185 \\
fighter & A &  2h21m &     1551 &        459 &       2.0678 \\
gargoyle & B &  0h29m &      835 &        225 &      10.5912 \\
gorilla & B  &  3h43m &     2609 &        956 &      12.3074 \\
explorer & B &  4h00m &     1700 &        808 &      17.2768 \\
engineer & A &  0h54m &      844 &        785 &       2.0379 \\
elf   & B &  1h37m &     4210 &        782 &       5.6241 \\
\bottomrule
\end{tabular}
}
\end{center}
\caption{Statistics on the models used in the paper: name, interaction time, number of strokes and camera movements and model size (estimated as bounding box diagonal).}
\label{tbl:models}
\end{table}


\newcommand{\cd}{\hspace{0in}}
\newcommand{\cw}{1.25in}
\newcommand{\tf}[1]{\textsf{#1}}
\begin{figure*}
    \begin{center}
    \begin{tabular}{@{}c@{}c@{\cd}c@{\cd}c@{\cd}c@{\cd}c@{}}
    &
    \tf{stroke length} & \tf{brush size} & \tf{stroke angles} & \tf{stroke pressure} & \tf{mesh distance}\\
    \rotatebox{90}{\hspace{0.4in}\tf{ogre}}&
    \includegraphics[width=\cw]{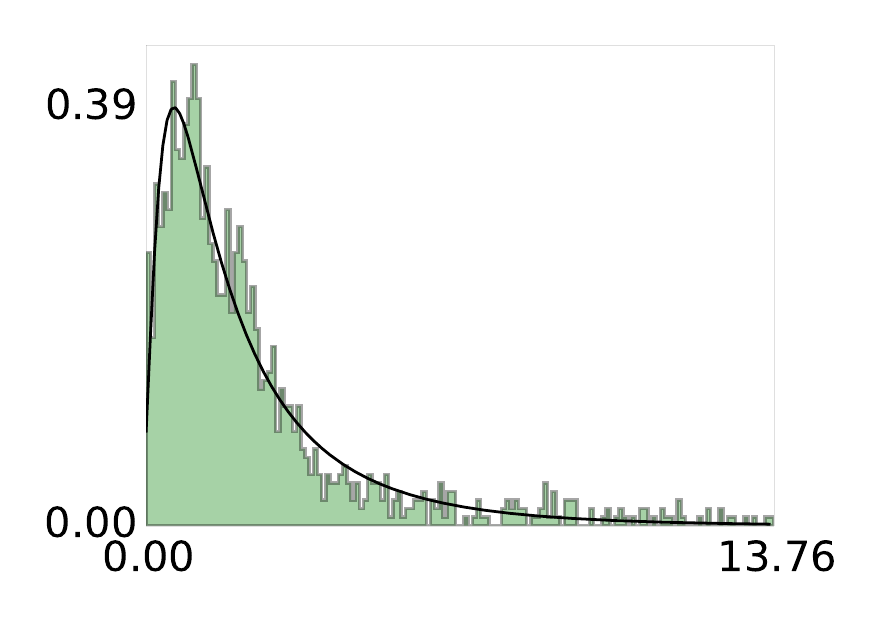}&
    \includegraphics[width=\cw]{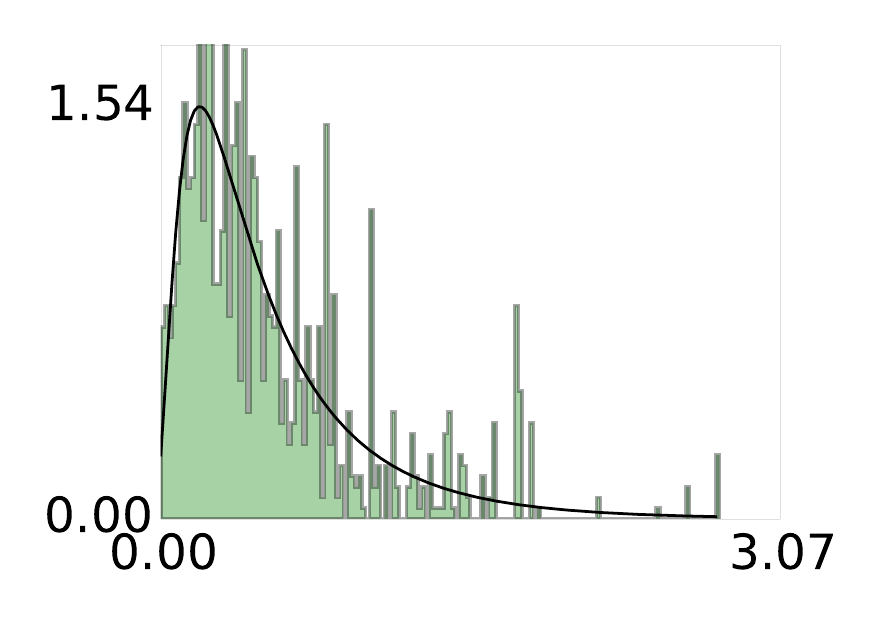}&
    \includegraphics[width=\cw]{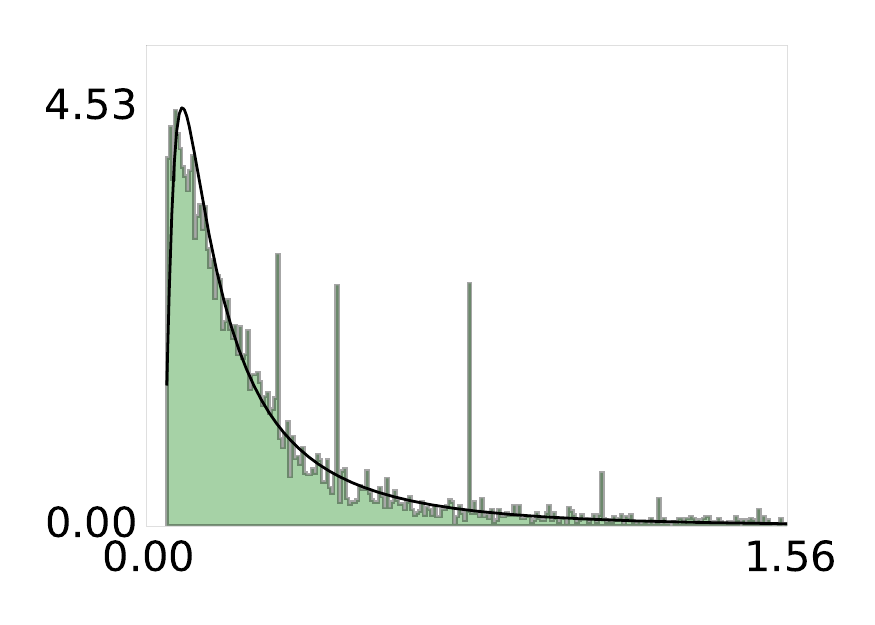}&
    \includegraphics[width=\cw]{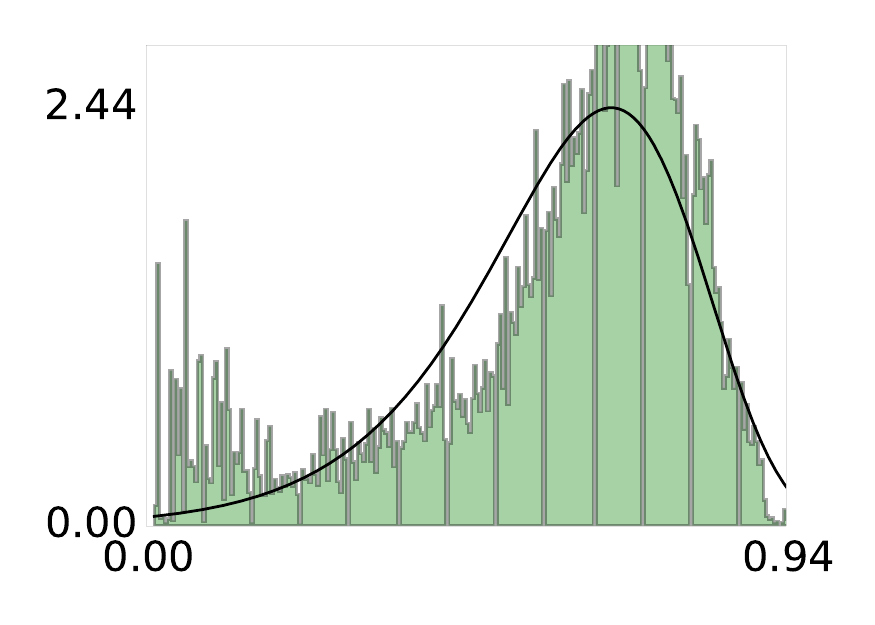}&
    \includegraphics[width=\cw]{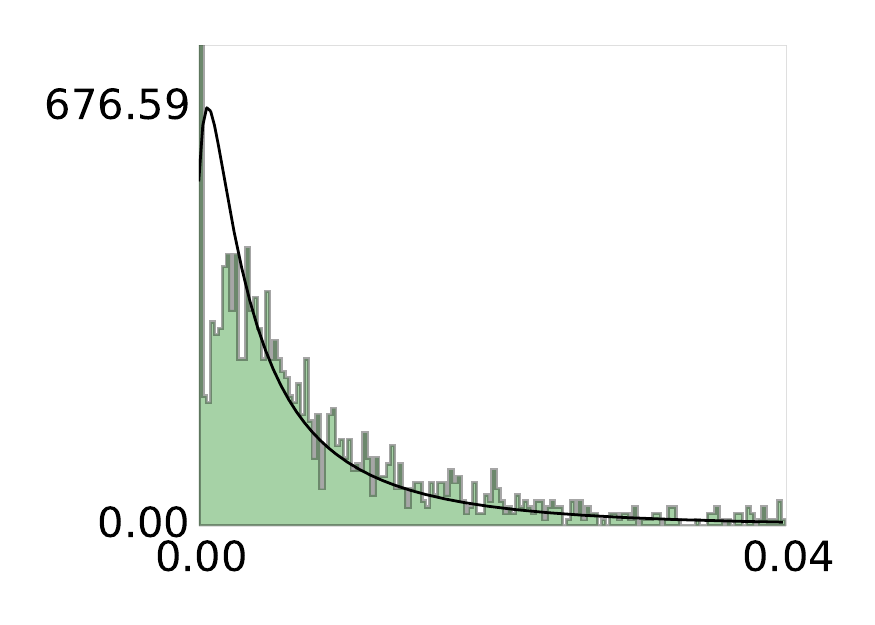}\\
    \rotatebox{90}{\hspace{0.4in}\tf{monster}}&
    \includegraphics[width=\cw]{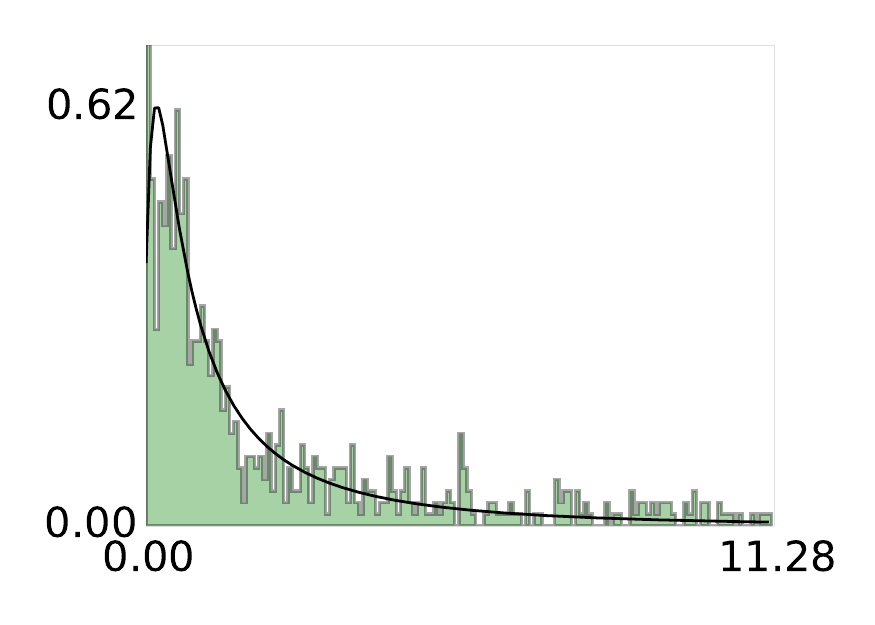}&
    \includegraphics[width=\cw]{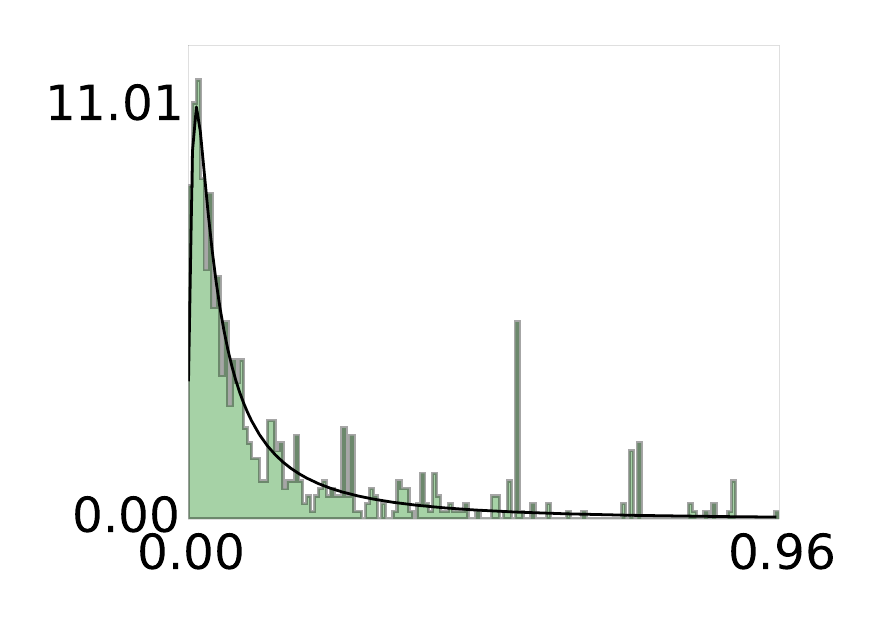}&
    \includegraphics[width=\cw]{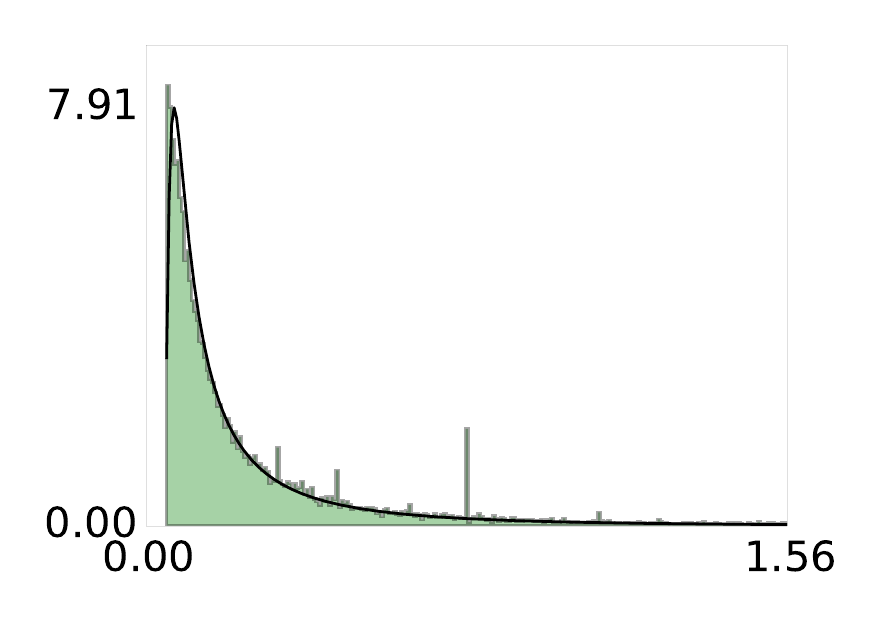}&
    \includegraphics[width=\cw]{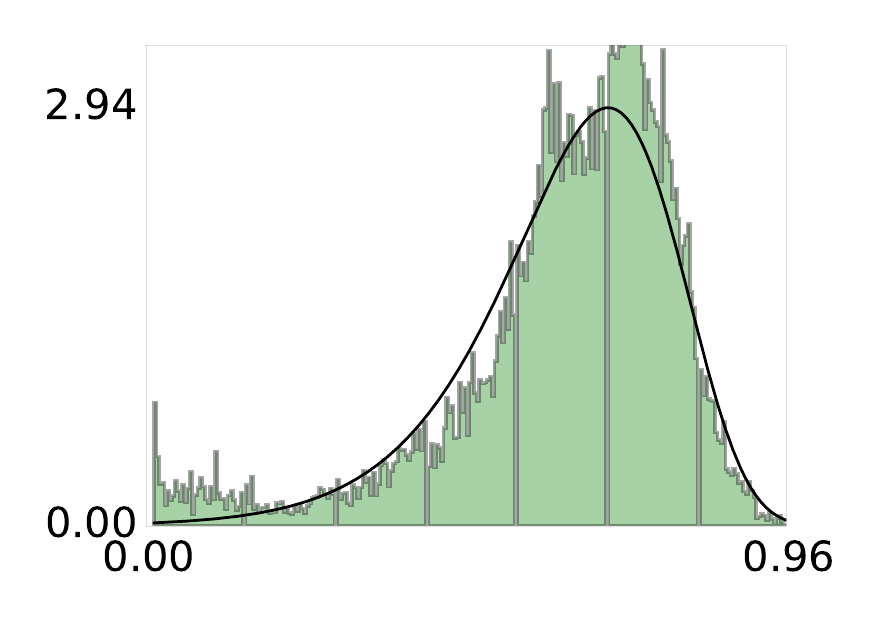}&
    \includegraphics[width=\cw]{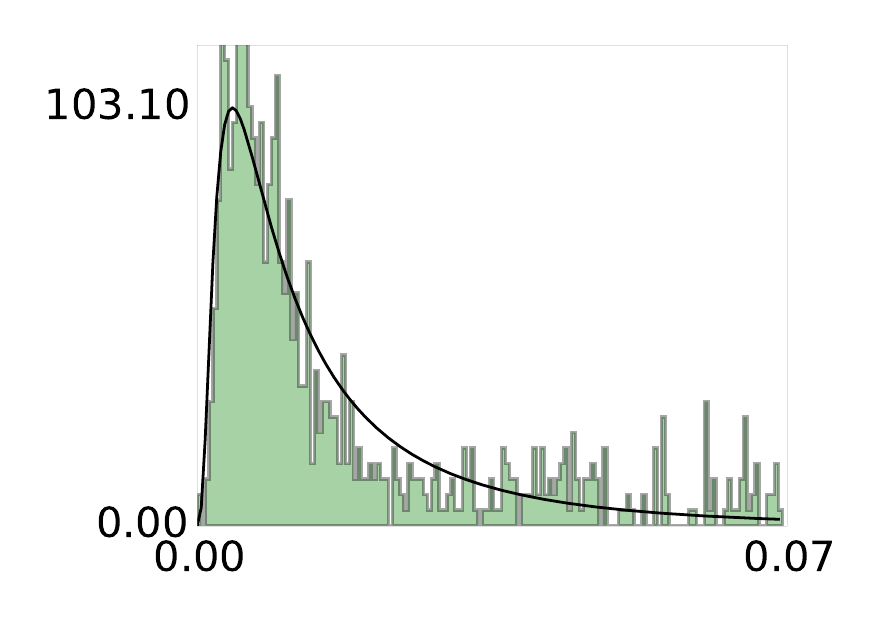}\\
    \rotatebox{90}{\hspace{0.4in}\tf{explorer}}&
    \includegraphics[width=\cw]{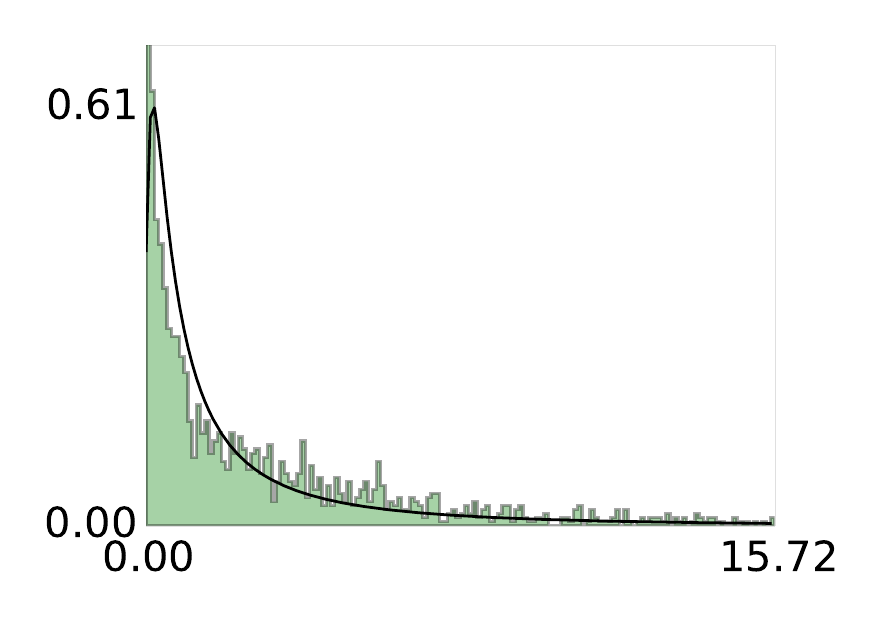}&
    \includegraphics[width=\cw]{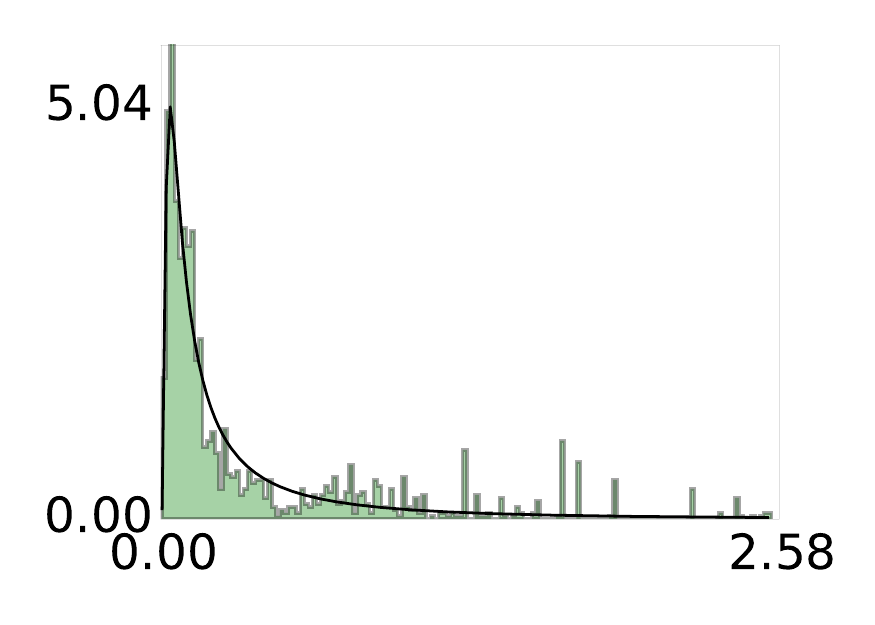}&
    \includegraphics[width=\cw]{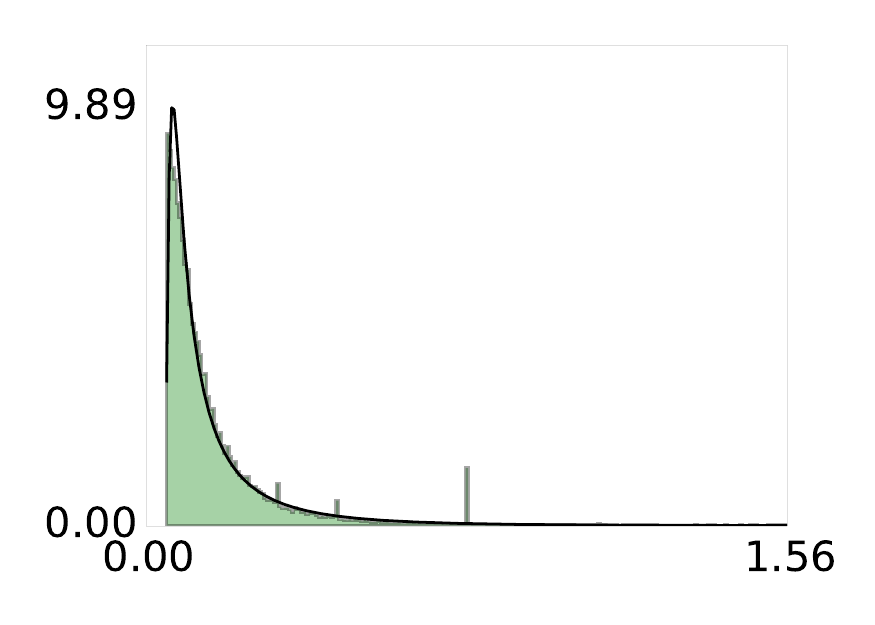}&
    \includegraphics[width=\cw]{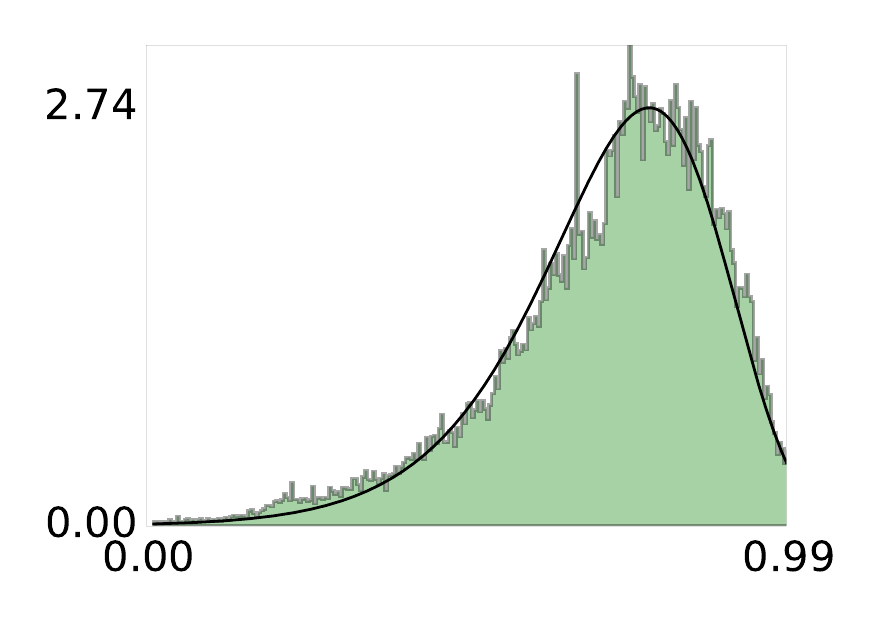}&
    \includegraphics[width=\cw]{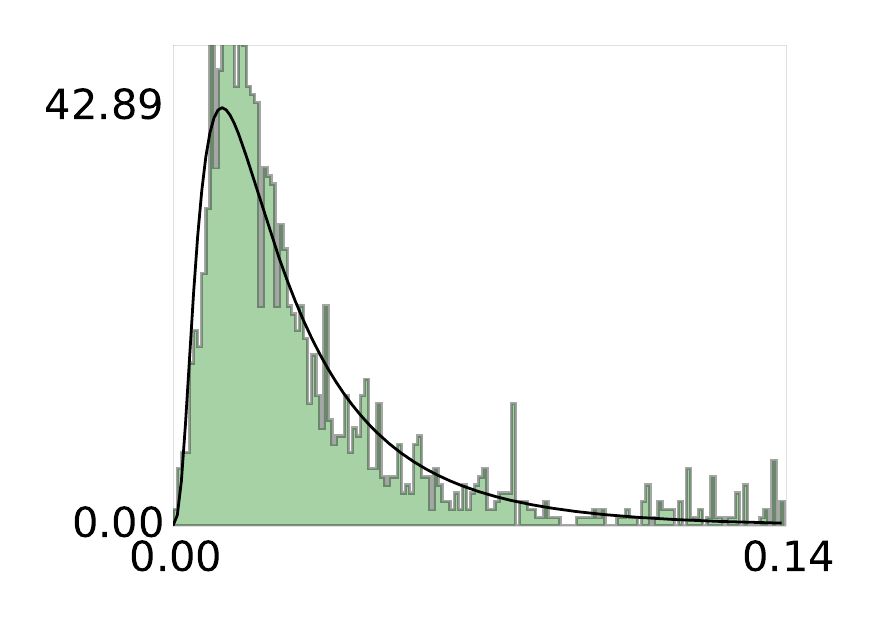}\\
    \end{tabular}
    \end{center}
    \caption{Histogram of analyzed parameters with fitted distributions for representative meshes. To optimize space, we show only values within the 95\% percentile.}
    \label{fig:overall}
\end{figure*}

\section{Analysis: Overall Trends}

We perform various types of analysis on stroke data and mesh distances. For each analysis, we qualitative illustrate trends using diagrams, and perform statistical estimation of model parameters and quality of fits. We include in the paper only representative diagrams, but provide all diagrams in supplemental. \tbl{tbl:overall} summarizes the findings discussed in this section, shown in \fig{fig:overall}. We analyze features separately since we did not find significant correlation between them, but we included scatterplots for all the brush attributes analyzed, in supplemental.

\parasection{Brush Type}
For all models in the paper, on-surface strokes account for the majority of interactions, with no significant difference between model types. We believe that this can be easily interpreted considering that free form strokes are used for extrusions and freeform deformations, two global operations that are rare. Furthermore, the statistics of on-surface and freeform strokes are close enough that we analyze them together and only highlight differences when present.

\parasection{Methodology}
For all brush attributes and mesh differences, we determine the theoretical distribution that best  describe the data by choosing the best fitting model. We report mean or standard deviation of the process to quantitatively characterize it. We consider several well-known distributions (Normal, Log-normal, Student's T, Cauchy, Inverse gaussian, as well as other main theoretical distributions). The parameters of each distribution are fitted by maximum likelihood estimation \cite{paramstats}, that maximizes the log-likelihood of the input given the set of parameters. We measure the goodness of fit with $\chi^2$ tests \cite{chi-square}. We select the best distribution by picking the one with highest $\chi^2$ amongst the ones with high $p$-values ($>0.05$), minimizing distribution complexity (i.e.: number of parameters).

\parasection{Stroke Length}
The distributions of stroke lengths, shown in \fig{fig:overall} are peaked near zero with long tails. This indicates that artists work mostly with quick and short strokes, but perform a good number of longer ones as well. For all models, stroke length is described best by an inverse gaussian distribution. The inverse gaussian, as a member of the exponential distributions' family, is well suited for extremely peaked data, also allowing for extreme values. The average stroke length differs for each model, since models have arbitrary scale. We observed that the mean length is related to model features: eyes and mouth's width for heads, neck length and diameter for busts and arms and legs for full-bodies. The maximum stroke length exceed the model diagonal. This suggest that artists perform long straight strokes to adjust specific features, but also stroke back and forth on the same area, mainly for smoothing or texturing purposes. We will consider these two behaviours later.

\parasection{Brush Size}
The distributions of brush sizes, shown in \fig{fig:overall} have similar shape to those of strokes' length, that can be described well with an inverse gaussian distribution. Similarly to the length, artists work the majority of their time with small-sized brushes. In this case though, the tails have a less homogeneous descending trend, with smaller peaks along the whole tail itself. This latter observation can be explained if we consider that artists usually work for relatively long time spans at fixed camera and 2D brush size. The resulting brush size projected over the model are thus relatively similar, forming small peaks in the distributions. The average size is related to models' features, but at a smaller scale compared to brush lengths: wrinkles and veins for the heads, eyes and ears for busts, hands and shoulders for full-body. The maximum size can reach the dimension of the whole model, and it usually happens with the use on freeform brushes when adjusting the proportion and the main structure of the whole model.

\parasection{Stroke Angles}
A useful feature to describe the stroking behaviors is the average angle between subsequent segments in a stroke ($\alpha_{12} = \arccos(seg_1, seg_2)$). We compute the angle projected in the camera plane since this is aligned with the mouse movements. In this case the fitting process was less obvious, as these distributions show (\fig{fig:overall}) two main peaks at $0$ and $\pi$. A deeper analysis showed that most of the strokes with broader angles were done with freeform brushes, demonstrating how artists favor long straight strokes for this kind of brushes. This is in opposition for what we observed analyzing angles for on-surface brushes. These distributions, which also in this case resulted to be better fitted by inverse gaussians, peak at zero, with long tails that only show a slow increasing trend towards the end, although only accounting for less than 5\% of the whole distribution. This indicates a strong preference for stroking back-and-forth in this kind of brushes, with a smaller percentage of long straight strokes.

\parasection{Stroke Pressure}
Strokes' pressure distributions, shown in \fig{fig:overall} are described well by gaussians, but with mean and standard deviation that differ significantly between on-surface and freeform brushes. On-surface brushes are used with a higher pressure mean and variance, demonstrated also by an higher index of dispersion on the distribution ($iod = 0.119$). On the other hand, freeform strokes are performed with less pressure, ranging in a statistically smaller spectrum of values ($iod = 0.051$). This supports the above observations that on-surface brushes are used for longer but detailed strokes, while freeform brushes are for quick but large adjustments.

\parasection{Mesh distances}
Hausdorff distances were fitted similarly to strokes' length and brush size, using inverse gaussian distributions. We don't supply any further interpretation for this data, as we think that the similar distributions' behaviour supports the correlation measures presented in the previous section, thus reinforcing the conclusions presented above for those two attributes. 

\parasection{Interpretation}
Before running the experiment, we expected lengths' and sizes' distributions to be gaussians, where artists naturally pick comfortable-to-perform strokes and work by repeating them throughout the model. The data supports a different interpretation. Most strokes focus on adjusting specific model features, where artists pick stroke lengths and brush sizes appropriate to the features' sizes. This also partially explains the long tails, supporting the use of longer strokes that cover whole features (e.g.: eye vs. leg in a full body model). Artists also use quick taps, of length near zero, to precisely refine small features with on-surface brushes, or adjust the main proportion of the model with large freeform brushes, i.e. moving the whole head or an arm. Artists often perform back-and-forth strokes on the same area with on-surface brushes, usually for smoothing or texturing purposes. This suggest that to precisely control surface features, repeating simple strokes is easier than precisely configuring brush parameters or controlling tablet pressure. 

\begin{table*}
\begin{center}
\scalebox{\tblscale}{
\begin{tabular}{lrrrrrrrrrrrr}
\toprule
\multicolumn{1}{c}{\textbf{}} & \multicolumn{2}{c}{\textbf{Length}} & \multicolumn{2}{c}{\textbf{Size}} & \multicolumn{2}{c}{\textbf{2D angles}} & \multicolumn{2}{c}{\textbf{Hausdorff}} & \multicolumn{2}{c}{\textbf{Pressure}} & \multicolumn{2}{c}{\textbf{Density}} \\
\multicolumn{1}{c}{\textbf{}} & \multicolumn{2}{c}{Inv. Gaussian} & \multicolumn{2}{c}{Inv. Gaussian} & \multicolumn{2}{c}{Inv. Gaussian} & \multicolumn{2}{c}{Inv. Gaussian} & \multicolumn{2}{c}{Gaussian}  & \multicolumn{2}{c}{Inv. Gaussian} \\
\multicolumn{1}{c}{{}} & \multicolumn{1}{c}{Avg} & \multicolumn{1}{c}{Std}  & \multicolumn{1}{c}{Avg} & \multicolumn{1}{c}{Std}  & \multicolumn{1}{c}{Avg} & \multicolumn{1}{c}{Std}  & \multicolumn{1}{c}{Avg} & \multicolumn{1}{c}{Std}  & \multicolumn{1}{c}{Avg} & \multicolumn{1}{c}{Std}  & \multicolumn{1}{c}{Avg} & \multicolumn{1}{c}{Std} \\
\midrule
alien    & 1.0472 & 1.6566 & 0.1028 & 0.0951 & 0.2632 & 0.6206 & 0.0007 & 0.0023 & 0.4429 & 0.2123 & 0.0157 & 0.0143 \\
elder    & 1.7489 & 2.2930 & 0.1915 & 0.1267 & 0.2321 & 0.5691 & 0.0012 & 0.0012 & 0.4008 & 0.1927 & 0.0392 & 0.0325 \\
ogre     & 3.3956 & 4.8483 & 0.5356 & 0.4619 & 0.1909 & 0.6191 & 0.0068 & 0.0098 & 0.5968 & 0.1916 & 0.0931 & 0.0777 \\
merman   & 0.7547 & 0.9524 & 0.0744 & 0.1026 & 0.1683 & 0.5192 & 0.0016 & 0.0029 & 0.5817 & 0.2234 & 0.0182 & 0.0167 \\
man      & 0.7886 & 1.4821 & 0.0411 & 0.0375 & 0.2563 & 0.5812 & 0.0007 & 0.0012 & 0.4986 & 0.2199 & 0.0091 & 0.0114 \\
monster  & 2.9010 & 5.2696 & 0.1525 & 0.2602 & 0.2404 & 0.6228 & 0.0173 & 0.0190 & 0.6227 & 0.1600 & 0.0309 & 0.0239 \\
sage     & 0.7832 & 1.4256 & 0.0906 & 0.0951 & 0.1708 & 0.5702 & 0.0009 & 0.0027 & 0.6465 & 0.1758 & 0.0121 & 0.0103 \\
fighter  & 0.2588 & 0.3811 & 0.0154 & 0.0078 & 0.2556 & 0.6502 & 0.0004 & 0.0003 & 0.5530 & 0.2061 & 0.0040 & 0.0023 \\
gargoyle & 2.0315 & 2.6720 & 0.2425 & 0.1592 & 0.1979 & 0.5715 & 0.0062 & 0.0123 & 0.6596 & 0.1490 & 0.0499 & 0.0395 \\
gorilla  & 3.4470 & 6.7520 & 0.5274 & 0.3506 & 0.2777 & 0.6792 & 0.0126 & 0.0083 & 0.5894 & 0.1943 & 0.0340 & 0.0247 \\
explorer & 3.9406 & 8.5568 & 0.3808 & 0.6716 & 0.2486 & 0.6567 & 0.0282 & 0.0254 & 0.6930 & 0.1821 & 0.0253 & 0.0224 \\
engineer & 0.7763 & 1.1086 & 0.0184 & 0.0266 & 0.2282 & 0.6596 & 0.0008 & 0.0012 & 0.6411 & 0.1800 & 0.0046 & 0.0040 \\
elf      & 0.4032 & 1.4723 & 0.1774 & 0.1246 & 0.4386 & 0.7797 & 0.0082 & 0.0146 & 0.4354 & 0.2185 & 0.0163 & 0.0149 \\
\bottomrule
\end{tabular}
}
\end{center}
\caption{Summary statistics of parameter distributions.}
\label{tbl:overall}
\end{table*}


\section{Analysis: Temporal Trends}

After analyzing global trends, we focus on their temporal aspects, by considering brush attributes and mesh differences over time as if they were time series, with the temporal progression given by the sequence of snapshots. \tbl{tbl:temporal} summarizes the results of our analysis.

\parasection{Exponential Smoothing}
To gain insights on temporal behaviors, we perform exponential smoothing to remove noise from the data \cite{exposmooth}. We chose exponential smoothing over moving average since the former can provide a finer approximation of the series, giving more information on how the values are changing timely, that would instead be smoothed out by a simple moving average. \fig{fig:temporal} shows a representative diagram, while the remaining are included in supplemental. The diagrams show that neither brush attributes nor mesh differences have visible trends other than showing a certain number of ``bursts'', whose position in time has no significant periodicity. An autocorrelation analysis confirms the lack of periodicity. For brush length, peaks correspond to times where the artist switches to work on new model parts, performing adjustments over large areas of it (e.g.: adding texture or smoothing the whole upper body or head). For brush size, peaks correspond to freeform adjustments of large areas. These peaks are followed by smaller adjustments over the same areas.

\parasection{ARMA Modeling}
The observations made above can be summarized in a temporal behavior that has no apparent trends, but instead repeated shocks without periodicity. We formalize the observations by modeling the temporal trends with an \textit{autoregressive moving average} (ARMA) process \cite{arma}, following the Box-Jenkins method \cite{boxjenkins}. First, we validate stationarity by visual comparison of autocorrelation plots with ARMA's theoretical ones and by performing an augmented Dickey-Fuller test for unit root detection (test results are shown in \tbl{tbl:temporal}). Second, we estimate the model's parameters $p$ (order of the \textit{autoregressive} part of the model) and $q$ (order of the \textit{moving average} part of the model) with maximum likelihood estimation by optimizing the Akaike information criterion of the model given our data \cite{AICfitting1}. Third, the goodness of fit was established analyzing the autocorrelation plot of the residual of the fitted values, performing a Ljung-Box Q test on them. \tbl{tbl:temporal} summarizes values of this procedure. As can be seen, ARMA fits our data very well, corroborating our qualitative observations.

\parasection{Hidden Markov Model fitting}
To better understand this bursting behavior, we fitted an Hidden Markov Model on the time series. We treated the bursts as one of the states in which the artist could work (e.g.: editing sparsely along the whole model versus focusing on a single area). The parameters of the HMM were fitted using an expectation-maximization approach, iteratively increasing the number of states of the model at each run. The metric used to evaluate the goodness of fit is the Bayesian Information Criteria: the fitting process was stopped when the BIC value of the new estimated model didn't suggest any significant improvement from the previous one. A representative example of the trained HMMs is given in \fig{fig:temporal}. In the estimated models, all states are characterized with higher probability of remaining in the same state, rather than transition to another one. States differ mainly in variance and transition probabilities distribution, where states with higher variance have also less probability to remain in the same state.

\begin{figure*}
    \begin{center}
    \begin{tabular}{@{}c@{}c@{}}
    \textsf{stroke length} & \textsf{brush size} \\
    \includegraphics[width=\diagramwidth]{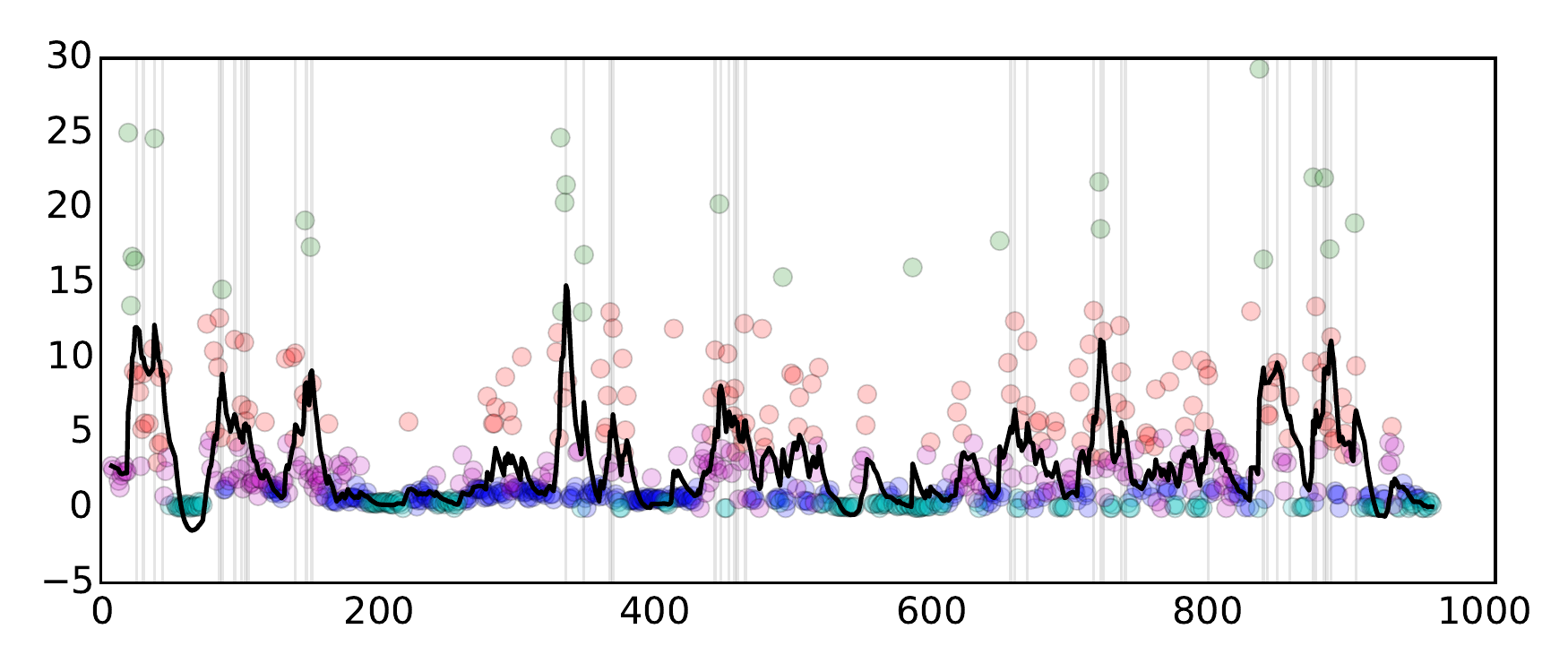} & \includegraphics[width=\diagramwidth]{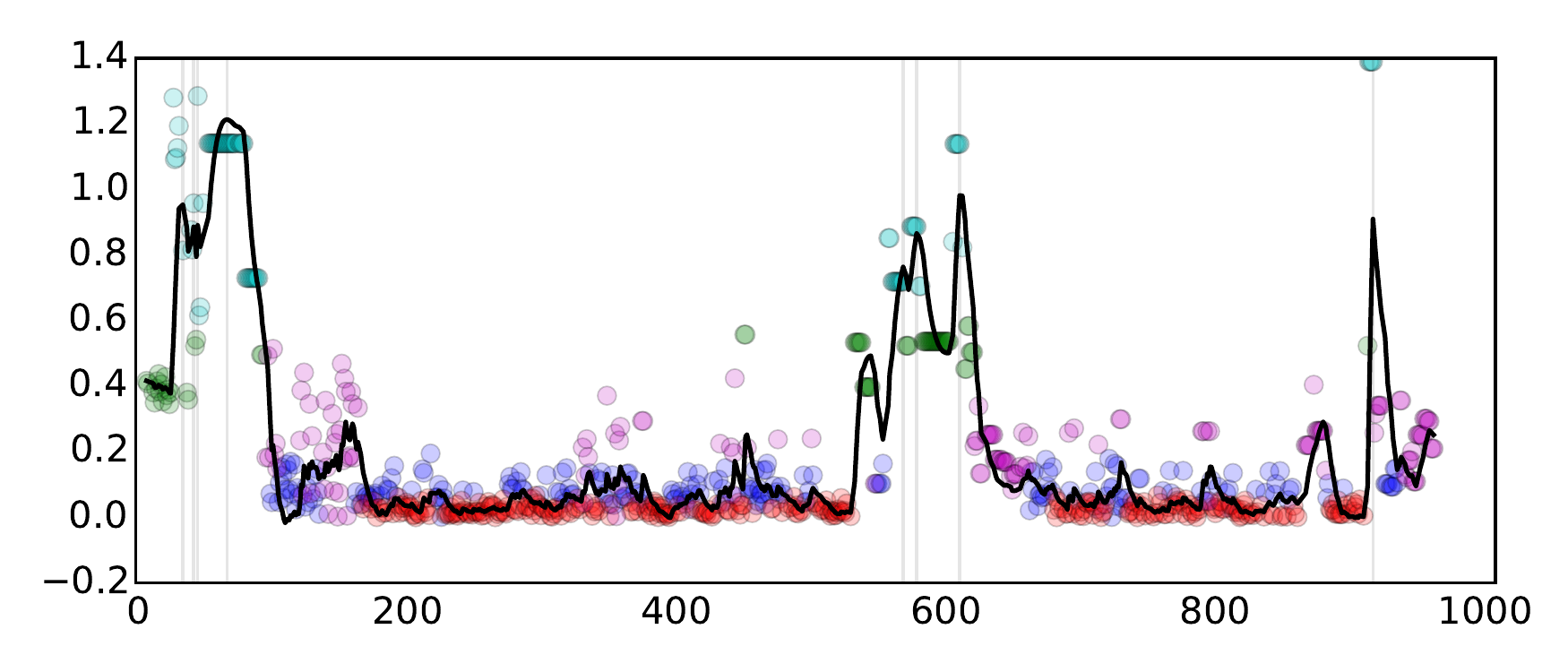} \\
    \end{tabular}
    \end{center}
    \caption{Stroke length and brush size for \emph{monster} model plotted over time. Exponential smoothing is performed to reduce noise. Parameter values are colored according to their hidden Markov state. We highlight relevant peaks with vertical lines.}
    \label{fig:temporal}
\end{figure*}

\parasection{Interpretation}
Before running this experiment, we expected artists to work in a coarse-to-fine manner, starting by  roughly defining the whole model and proceeding with gradual refinements of increasingly smaller magnitude. Our analysis demonstrates a completely different workflow. Artists mainly work in bursts of activity, alternating periods of refinement and addition of small features to bigger and broader changes. Also, they don't complete a single part of the model before working on another one, but instead going back and forth over the whole model during the bursts, often returning on the same parts.

\begin{table}
\begin{center}
\scalebox{\tblscale}{
\begin{tabular}{lrrrrrrr}
\toprule
{} & \textbf{ADF test}  & \textbf{p} & \textbf{q} & \textbf{Q test} & \textbf{Peaks} \\
\midrule
alien    & -14.08  &  2 &  1 &  64.9 &     15 \\
elder    & -16.22 &  4 &  1 &  282.13  &     15 \\
ogre     & -14.34  &  1 &  1 &   58.95  &     12 \\
merman   & -18.02  &  2 &  1 &   97.76  &     12 \\
man      & -12.62  &  3 &  2 &  162.36  &     13 \\
monster  &  -9.17  &  2 &  2 &   31.79  &      7 \\
sage     & -11.92  &  2 &  1 &   63.29  &     11 \\
fighter  & -12.9  &  4 &  2 &   94.64  &      7 \\
gargoyle &  -8.94  &  2 &  1 &   64.37  &      5 \\
gorilla  & -15.59  &  4 &  2 &   94.23 &     18 \\
explorer & -15.74  &  4 &  2 &   72.74  &      8 \\
engineer &  -8.31  &  3 &  1 &   30.59  &      3 \\
elf      & -17.06  &  4 &  1 &   76.92 &     35 \\
\bottomrule
\end{tabular}
}
\end{center}
\caption{Summary statistics of temporal distributions. Refer to the text for metric definitions.}
\label{tbl:temporal}
\end{table}


\begin{figure*}
    \begin{center}
    \includegraphics[width=\meshwidth]{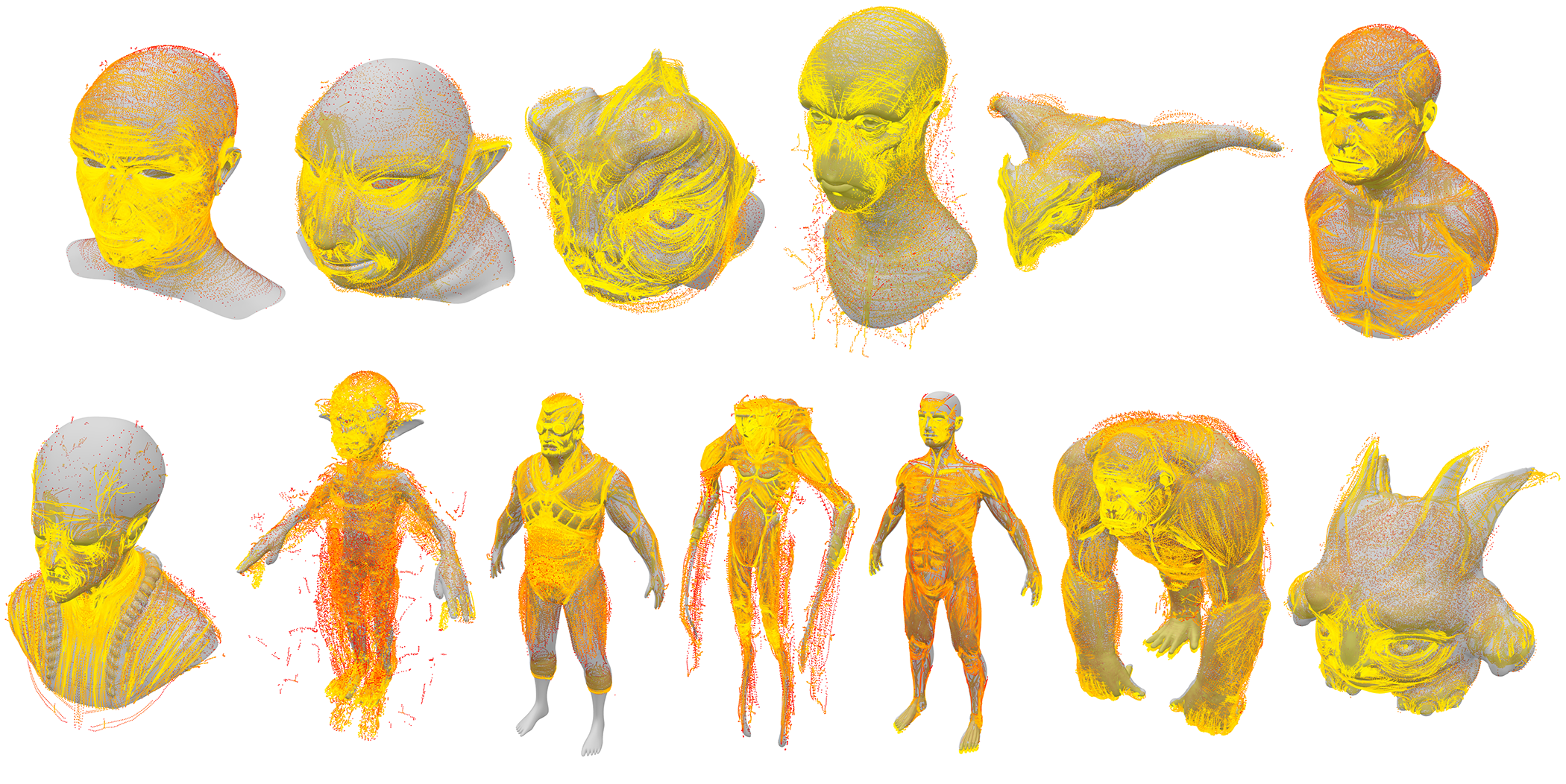}
    \end{center}
    \caption{Stroke density rendered over the mesh as reference. Red to yellow indicate low to high density.}
    \label{fig:density}
\end{figure*}

\begin{figure}
    \begin{center}
    \rotatebox{90}{\hspace{0.4in}\tf{explorer}}
    \includegraphics[width=\cw]{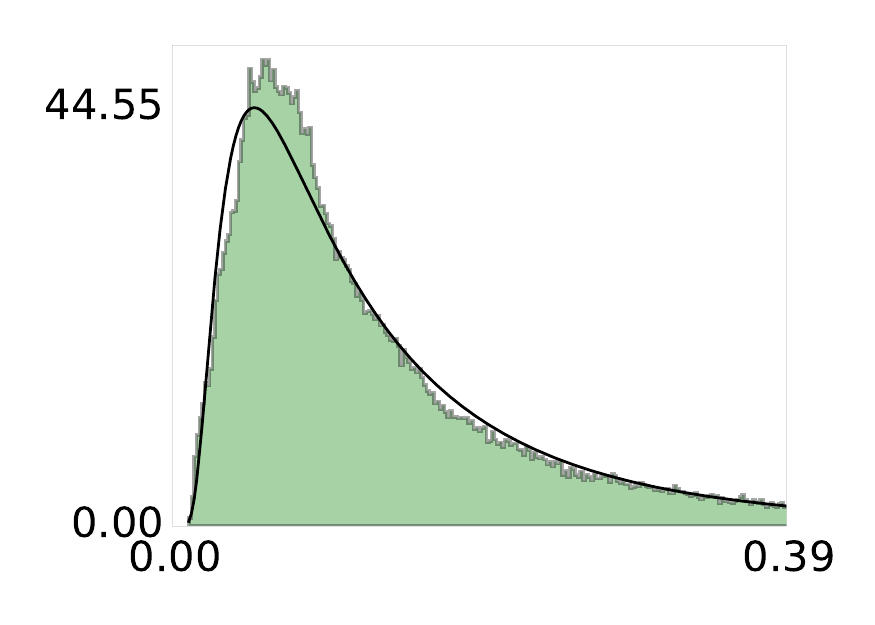}
    \rotatebox{90}{\hspace{0.4in}\tf{elder}}
    \includegraphics[width=\cw]{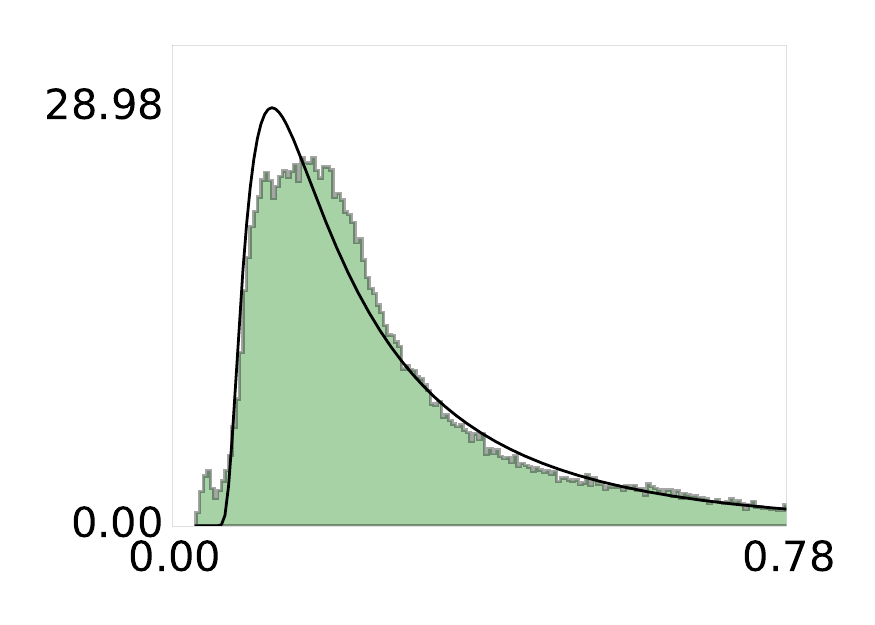}
    \end{center}
    \caption{Histogram of stroke densities estimated as average distance of 16 nearest neighbors. To optimize space, we show only values within the 95\% percentile.}
    \label{fig:density2}
\end{figure}

\section{Analysis: Spatial Trends}

After investigating the temporal characteristics of sculpting, we focus on determining how brush attributes and mesh differences behaved spatially, by checking whether these attributes are related to the stroke polyline spatial locations.

\parasection{Heat maps}
For every model, we estimate the density of the point cloud generated by the vertices of the polylines of all the strokes applied to it. For each point, we compute the average distance of its 16 nearest neighbors as our density estimate. \fig{fig:density} show mesh densities, and \fig{fig:density2} representative distributions. From these figures it is clear that the artists focus on the some areas more intensely than on others: eyes and mouth for faces and busts, head and torso for full bodies.

\parasection{Density distribution}
We then fitted the distributions of the computed densities with the same approach we used for the brush attributes in the overall analysis. Also in this case, we fitted using a inverse gaussian distribution. The peaks near zero supports the idea that most of the artists' edits are localized in just a set of areas. Interestingly, some of the distribution displayed a second peak nearer to zero. This bimodal trend can be interpreted as the difference of scale between the different features of the model (i.e.: the size of the eye versus the size of an arm).

\parasection{Spatial correlation}
The former analysis is purely spatial, in the sense that takes into account just the information on the position where the brush was applied. Since it shows that artists work more frequently on some parts of the models rather than others, we then analyzed how brush attributes vary spatially. More precisely we want to determine whether artists stroking behavior is specific to the location on the model. We perform a spatial autocorrelation test on the data, analyzing each attribute individually \cite{spatial}, that correlates attributes with multidimensional spatial locations. We used Moran's $I$ as in indicator. Values of $I$ range from $+1$ (perfect correlation, high values of the attribute are closer, same thing for lower values) to $-1$ (perfect dispersion), while a $0$ value indicates no correlation, i.e. a random spatial pattern. In this case, for computational reasons, we reduced the point cloud of all the points, to the point cloud of the centroids of all brush strokes, transferring to them all the attributes of the original brush stroke. Stroke length and brush size have a positive autocorrelation with $p$-value $<0.05$ for all models. All the autocorrelation values are included in the supplemental material, and \fig{fig:density} shows some of the point clouds, where each is point is colored according to the values of the attribute analyzed.

\parasection{Interpretation}
This analysis leads to a simple interpretation. While one would expect that artists focus more on important or detailed model areas, such focus is less strong than expected. The spatial correlation suggests that different techniques are applied to different locations, confirming that artists change stroking behavior to adapt to the specific feature they are working on.


\begin{figure*}[tb]
    \begin{center}
    \includegraphics[width=\meshwidth]{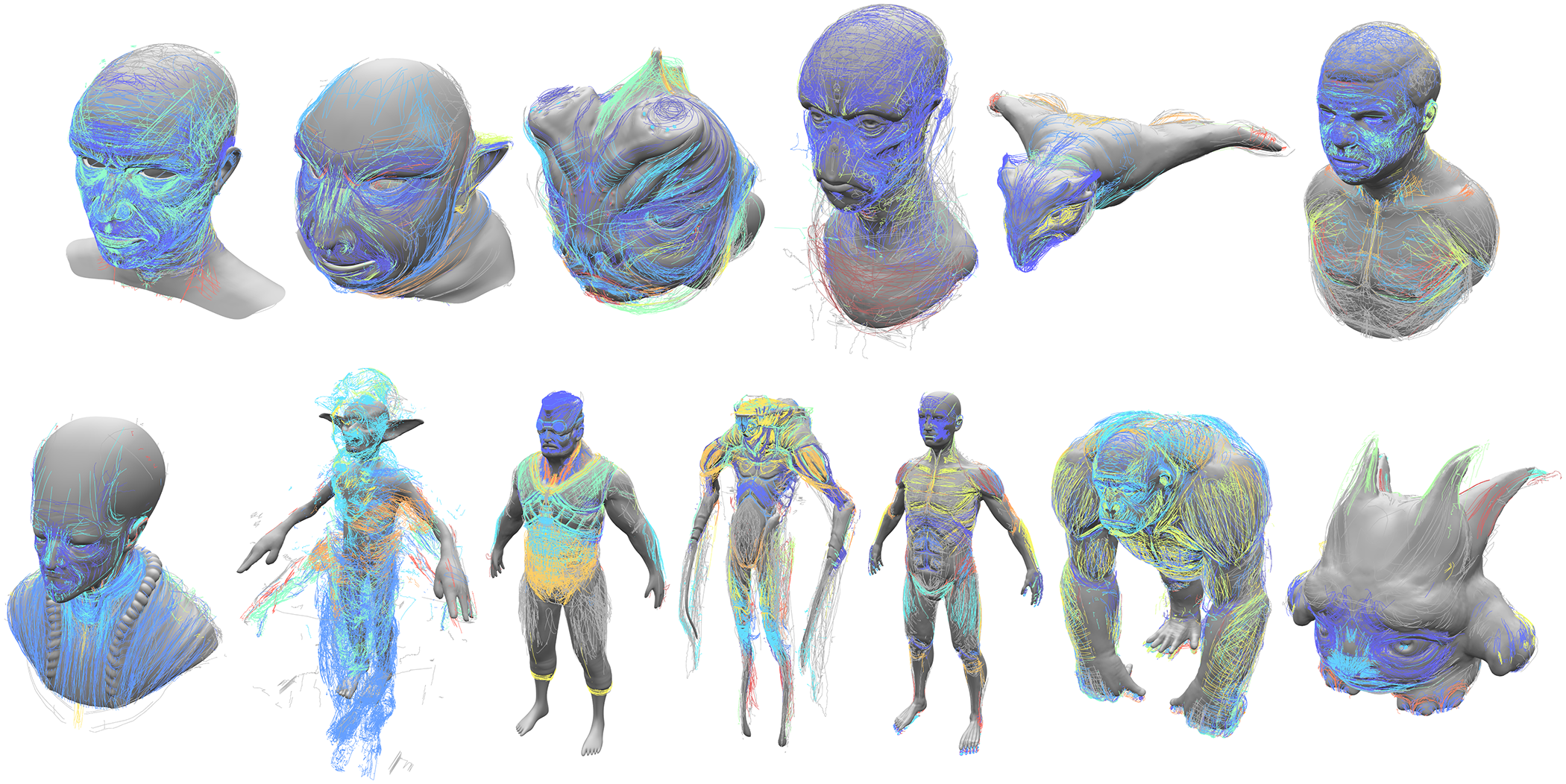}\\
    \end{center}
    \caption{Spatio-temporal clusters of strokes. Colors indicate cluster assignment.}
    \label{fig:spatiotemporal}
\end{figure*}

\section{Analysis: Spatial-temporal Analysis}

The temporal analysis shows that artists don't proceed in corse-to-fine manner, but alternate longer phases of small localized adjustments with phases of broader edits involving larger parts of the model. The spatial analysis shows artists' tendency to concentrate on selected areas of the model. We now investigate the spatio-temporal relationship between edits.

\parasection{Spatio-Temporal Clustering}
To explore this behavior, we cluster strokes taking into account spatial (strokes' centroids), non-spatial (path length, brush size, and mesh difference) and temporal (stroke time) information. To properly account for all information above, we used the ST-DBSCAN algorithm \cite{stdbscan}, a density-based clustering method. The density-based component of the algorithm allows clusters to have any shape spatially. Temporal distance is accounted for with a moving window. All other properties are treated as a vector of floating point attributes.  This kind of clustering is particularly useful in our context, because it can adapt well on non-flat geometries and uneven cluster sizes. Our data has these characteristics since strokes follow models' shapes.

\begin{figure*}[tb]
    \begin{center}
    \begin{tabular}{@{}c@{}c@{}}
    \textsf{stroke length} & \textsf{brush size} \\
    \includegraphics[width=\diagramwidth]{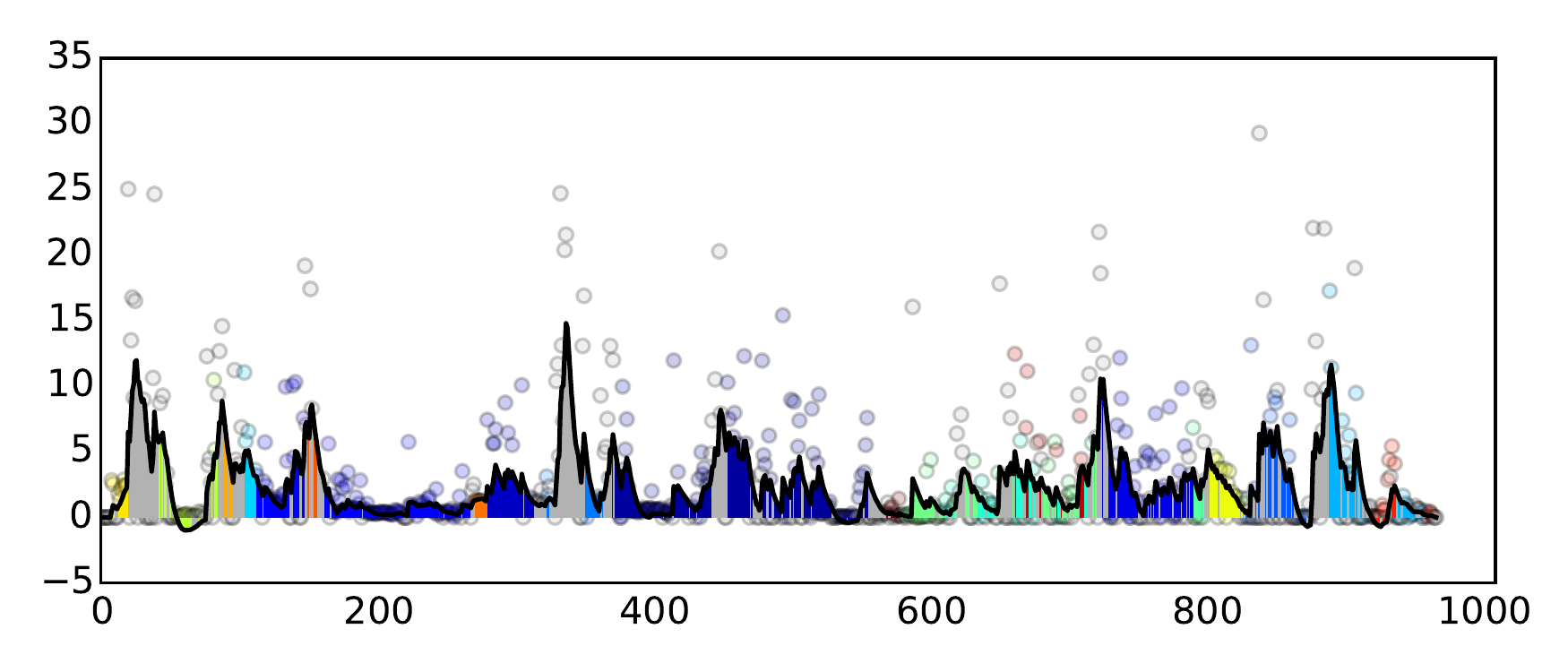} & \includegraphics[width=\diagramwidth]{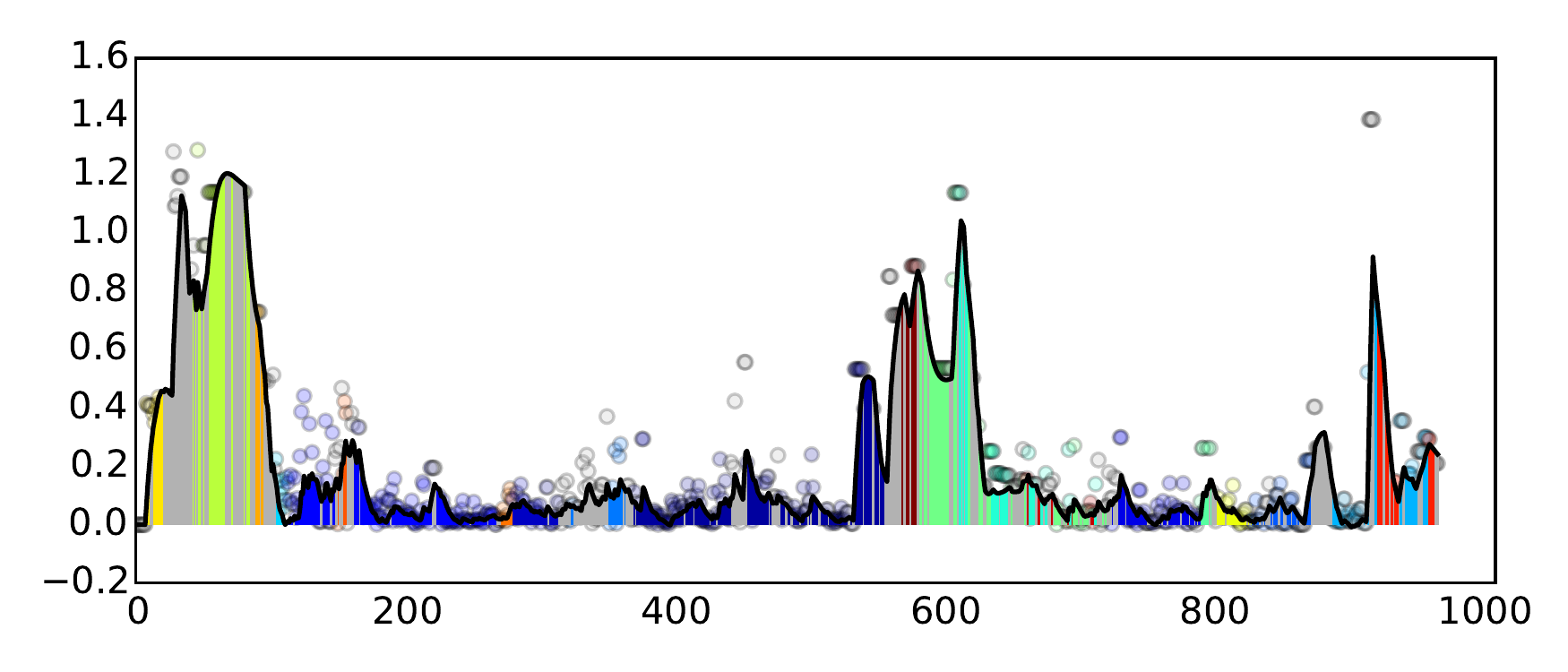} \\
    \end{tabular}
    \end{center}
    \caption{Exponential smoothing of clustered stroke length and brush size for the \emph{monster} model. Colors indicate cluster assignment.}
    \label{fig:spatiotemporal2}
\end{figure*}

\fig{fig:spatiotemporal} shows clustered strokes and \fig{fig:spatiotemporal2} shows representative results of the clustering properties plotted over time. Clusters overlap on the same mesh regions, indicating that artists work on the same regions at different times. Each time, brush properties stay similar within the activity burst, and artists stay on the region for a while before switching to another one. From the temporal plots we can see that stroke attributes are temporally related to the clusters.

\parasection{Interpretation}
The latter three analyses provide us with a clear picture about artists workflow. During sculpting, artists start focusing on some areas, dedicating them a good amount of work, then moving to a phase of adjustments and eventually coming back to the same spots for finer edits.

\section{Discussion and Limitations}

\parasection{Methodology}
Throughout the paper we employed a variety of state-of-the-art statistical methods to study digital sculpting behavior in an unguided experiment. The large dataset used (roughly 25000 strokes/mesh snapshots) allows us to extract statistically significant trends even from ``free-form'' data. In particular, overall trends and spatio-temporal behaviors would not have been possible to measure with targeted experiments, since they bias artists and they generally use smaller datatsets.

\parasection{Results}
We summarize here the main findings of our analysis.
\begin{compactenum}
    \item A relatively small number of strokes is sufficient to create relatively detailed models.
    \item Brush parameters are changed rarely, indicating that artists prefer simple brushes used repeatedly than configuring complex brush precisely.
    \item Strokes length, brush size and mesh differences are described well by inverse gaussian distributions, made mostly of short strokes, but with a thick tail of longer strokes.
    \item The average of stroke length and brush size is related to model features, suggesting that artists change their stroke patterns to adapt their workflow to the model.
    \item Strokes are either very short for precise transformations, medium length and straight to follow a feature, or long with repeated back-and-forth movements for controlling surface ``texture''.
    \item Temporally, artists do not work in a coarse-to-fine fashion, but in bursts of activity on different parts of the mesh.
    \item Temporal behavior is stationary with no significant periodicity.
    \item Bursts can be modeled as states in a hidden Markov model where they are stable, in that artists work in these states for a while before transitioning to another one.
    \item Spatially, artists focus on some selected regions by dedicating different amounts of edits and by applying different techniques.
    \item Strokes' parameters are correlated with spatial location, proving that artists use different techniques on different parts of the mesh.
    \item Bursts can be modeled as spatio-temporal clusters, showing that artists return on the same model region multiple times to perform different techniques.
    \item Spatio-temporal clusters prove that a coarse-to-fine pattern in not present even locally.
\end{compactenum}

\parasection{Limitations: Experiment Size}
As with all user studies, limitations often depend directly on experiment size in that more questions can be answered by increasing the number of subjects and workflows. For this work, the main limitation of this type is that we do not perform inter-subject analysis. The simple motivation for this is data scale. The dataset in this experiment takes roughly 100 GB to store compressed. The most complex analysis we performed took several hours of computation per mesh. We estimate that roughly 100 times more data would be needed for a statistically significant inter-subject analysis, but only if models choices are not left to subjects. This break our first desiderata. For these reasons, we did not perform this analysis. While it is possible that the conclusions of this work and the introduced methodology can be used to design a targeted experiment for inter-subject analysis, it remains unclear how to avoid the bias introduce by guidance. So we leave this investigation to future work. Similarly, we do not consider novice users, mainly since it remains unclear that users can model detailed meshes without training. It would still be interesting though to see whether simplified sculpting interfaces can be useful to such users. For this though, we would suggest an experiment design similar to \cite{lightfields}.

\parasection{Limitations: Exploratory Analysis}
Throughout the paper we took the approach of interrogating data in an exploratory manner, rather than testing specific hypothesis of workflows characteristics. We tried the latter, but failed since we found that many hypothesis about artists behaviors made watching videos or discussed in the literature were not supported by the data. For example, coarse-to-fine trends are cited in most graphics literature. On the other hand, it is certainly useful to be able to answer very specific question doing hypothesis testing. For this very reason we release all our data. 

\parasection{Impact}
We believe that a scientific characterization of content creation tasks is necessary to advance in this topic, now that our field is more mature. This paper is just a first step in doing it, producing statistically valid results for a specific creation task, namely sculpting. The impact of our work is twofold. First, using this data current interface might be improved. For example, it is clear that highly configurable brushes are not particularly useful since artists consistently prefer to use simpler settings with more strokes, since these provide better control. Note though that this is the opposite trend of many packages now. Second, artists routinely change models proportion significantly after details have been created. This suggest the need for detail-preserving transformations that alter the mesh while maintain its ``texture'' intact. Today soft transformations deform details too significantly, leading to resculpting. More importantly though we believe that the main impact of this work is our methodology that can be applied to other content creation tasks, such as material painting, environment-map lighting, keyframe animation, etc. 


\section{Conclusions and Future Work}

In conclusion, we present a methodology to statistically characterize 3D content creation workflows and apply it to investigate digital sculpting. We analyze the creation of several meshes, both from scratch and from base meshes. We use statistical methods to extract trends in the underlying data and ensuring that such trends are significant. We plan to use a similar methodology for investigating other content creation tasks as future work.

\bibliographystyle{acmsiggraph}
\bibliography{sculptstat15}

\begin{thebibliography}{\protect\citename{Pfanzagl and Hamb{\"o}ker }1994}

\bibitem[\protect\citename{Angelidis et~al\mbox{.} }2006]{sweeper}
{\sc Angelidis, A., Wyvill, G., and Cani, M.-P.}
\newblock 2006.
\newblock Sweepers: Swept deformation defined by gesture.
\newblock {\em Graph. Models 68}, 1, 2--14.

\bibitem[\protect\citename{Bae et~al\mbox{.} }2008]{ilovesketch}
{\sc Bae, S.-H., Balakrishnan, R., and Singh, K.}
\newblock 2008.
\newblock Ilovesketch: As-natural-as-possible sketching system for creating 3d
  curve models.
\newblock In {\em UIST '08}.

\bibitem[\protect\citename{Berger et~al\mbox{.} }2013]{berger}
{\sc Berger, I., Shamir, A., Mahler, M., Carter, E., and Hodgins, J.}
\newblock 2013.
\newblock Style and abstraction in portrait sketching.
\newblock {\em ACM Trans. Graph. 32}, 4, 55:1--55:12.

\bibitem[\protect\citename{Birant and Kut }2007]{stdbscan}
{\sc Birant, D., and Kut, A.}
\newblock 2007.
\newblock St-dbscan: An algorithm for clustering spatial-temporal data.
\newblock {\em Data Knowl. Eng. 60}, 1, 208--221.

\bibitem[\protect\citename{Blender }2014]{blender}
{\sc Blender}, 2014.
\newblock Blender.
\newblock \url{http://www.blender.org/}.

\bibitem[\protect\citename{Box and Jenkins }1990]{boxjenkins}
{\sc Box, G. E.~P., and Jenkins, G.}
\newblock 1990.
\newblock {\em Time Series Analysis, Forecasting and Control}.
\newblock Holden-Day, Incorporated.

\bibitem[\protect\citename{Brockwell and Davis }1991]{AICfitting1}
{\sc Brockwell, P.~J., and Davis, R. A.~m.}
\newblock 1991.
\newblock {\em Time series : theory and methods}.
\newblock Springer series in statistics. Springer.

\bibitem[\protect\citename{Callahan et~al\mbox{.} }2006]{vistrails}
{\sc Callahan, S.~P., Freire, J., Santos, E., Scheidegger, C.~E., Silva, C.~T.,
  and Vo, H.~T.}
\newblock 2006.
\newblock Vistrails: Visualization meets data management.
\newblock In {\em SIGMOD '06}, SIGMOD '06, 745--747.

\bibitem[\protect\citename{Chen et~al\mbox{.} }2011]{nonlinear}
{\sc Chen, H.-T., Wei, L.-Y., and Chang, C.-F.}
\newblock 2011.
\newblock Nonlinear revision control for images.
\newblock {\em ACM Trans. Graph. 30}, 4, 105:1--105:10.

\bibitem[\protect\citename{Chen et~al\mbox{.} }2014]{3dauthoring}
{\sc Chen, H.-T., Grossman, T., Wei, L.-Y., Schmidt, R.~M., Hartmann, B.,
  Fitzmaurice, G., and Agrawala, M.}
\newblock 2014.
\newblock History assisted view authoring for 3d models.
\newblock In {\em SIGCHI '14}, SIGCHI '14, 2027--2036.

\bibitem[\protect\citename{Chernoff and Lehmann }1954]{chi-square}
{\sc Chernoff, H., and Lehmann, E.~L.}
\newblock 1954.
\newblock {\em Ann. Math. Statist. 25}, 3 (09), 579--586.

\bibitem[\protect\citename{Cignoni et~al\mbox{.} }1998]{metro}
{\sc Cignoni, P., Rocchini, C., and Scopigno, R.}
\newblock 1998.
\newblock Metro: Measuring error on simplified surfaces.
\newblock {\em Computer Graphics Forum 17}, 2, 167--174.

\bibitem[\protect\citename{Cole et~al\mbox{.} }2008]{drawlines}
{\sc Cole, F., Golovinskiy, A., Limpaecher, A., Barros, H.~S., Finkelstein, A.,
  Funkhouser, T., and Rusinkiewicz, S.}
\newblock 2008.
\newblock Where do people draw lines?
\newblock {\em ACM Trans. Graph. 27}, 3, 88:1--88:11.

\bibitem[\protect\citename{Denning and Pellacini }2013]{meshgit}
{\sc Denning, J.~D., and Pellacini, F.}
\newblock 2013.
\newblock Meshgit: Diffing and merging meshes for polygonal modeling.
\newblock {\em ACM Trans. Graph. 32}, 4, 35:1--35:10.

\bibitem[\protect\citename{Denning et~al\mbox{.} }2011]{meshflow}
{\sc Denning, J.~D., Kerr, W.~B., and Pellacini, F.}
\newblock 2011.
\newblock Meshflow: Interactive visualization of mesh construction sequences.
\newblock {\em ACM Trans. Graph. 30}, 4, 66:1--66:8.

\bibitem[\protect\citename{Dobo\v{s} and Steed }2012]{3ddiff}
{\sc Dobo\v{s}, J., and Steed, A.}
\newblock 2012.
\newblock 3d diff: An interactive approach to mesh differencing and conflict
  resolution.
\newblock In {\em SIGGRAPH Asia '12 Technical Briefs}, 20:1--20:4.

\bibitem[\protect\citename{Dobo\v{s} et~al\mbox{.} }2014]{3dtimeline}
{\sc Dobo\v{s}, J., Mitra, N.~J., and Steed, A.}
\newblock 2014.
\newblock 3d timeline: Reverse engineering of a part-based provenance from
  consecutive 3d models.
\newblock {\em Computer Graphics Forum\/}.

\bibitem[\protect\citename{Drucker et~al\mbox{.} }2006]{slides}
{\sc Drucker, S.~M., Petschnigg, G., and Agrawala, M.}
\newblock 2006.
\newblock Comparing and managing multiple versions of slide presentations.
\newblock In {\em UIST '06}, UIST '06, 47--56.

\bibitem[\protect\citename{Hu et~al\mbox{.} }2013]{inverse}
{\sc Hu, S.-M., Xu, K., Ma, L.-Q., Liu, B., Jiang, B.-Y., and Wang, J.}
\newblock 2013.
\newblock Inverse image editing: Recovering a semantic editing history from a
  before-and-after image pair.
\newblock {\em ACM Trans. Graph. 32}, 6, 194:1--194:11.

\bibitem[\protect\citename{Jarabo et~al\mbox{.} }2014]{lightfields}
{\sc Jarabo, A., Masia, B., Bousseau, A., Pellacini, F., and Gutierrez, D.}
\newblock 2014.
\newblock How do people edit light fields?
\newblock {\em ACM Trans. Graph. 33}, 4, 146:1--146:10.

\bibitem[\protect\citename{Karl{\'{i}}k et~al\mbox{.} }2014]{global}
{\sc Karl{\'{i}}k, O., R{\r{u}}{\v{z}}i{\v{c}}ka, M., Gassenbauer, V.,
  Pellacini, F., and K{\v{r}}iv{\'{a}}nek, J.}
\newblock 2014.
\newblock Toward evaluating the usefulness of global illumination for novices
  in lighting design tasks.
\newblock {\em IEEE TVCG 20\/}.

\bibitem[\protect\citename{Kerr and Pellacini }2009]{lighting}
{\sc Kerr, W.~B., and Pellacini, F.}
\newblock 2009.
\newblock Toward evaluating lighting design interface paradigms for novice
  users.
\newblock {\em ACM Trans. Graph. 28}, 3, 26:1--26:9.

\bibitem[\protect\citename{Kerr and Pellacini }2010]{materials}
{\sc Kerr, W.~B., and Pellacini, F.}
\newblock 2010.
\newblock Toward evaluating material design interface paradigms for novice
  users.
\newblock {\em ACM Trans. Graph. 29}, 4, 35:1--35:10.

\bibitem[\protect\citename{Kong et~al\mbox{.} }2012]{delta}
{\sc Kong, N., Grossman, T., Hartmann, B., Agrawala, M., and Fitzmaurice, G.}
\newblock 2012.
\newblock Delta: A tool for representing and comparing workflows.
\newblock In {\em SIGCHI '12}, CHI '12, 1027--1036.

\bibitem[\protect\citename{Moran }1950]{spatial}
{\sc Moran, P.}
\newblock 1950.
\newblock {Notes on continuous stochastic phenomena.}
\newblock {\em Biometrika 37\/}, 17--23.

\bibitem[\protect\citename{Mudbox }2014]{mudbox}
{\sc Mudbox}, 2014.
\newblock Mudbox.
\newblock \url{http://www.autodesk.com/products/mudbox/overview/}.

\bibitem[\protect\citename{Pfanzagl and Hamb{\"o}ker }1994]{paramstats}
{\sc Pfanzagl, J., and Hamb{\"o}ker, R.}
\newblock 1994.
\newblock {\em Parametric Statistical Theory}.
\newblock De Gruyter textbook. W. de Gruyter.

\bibitem[\protect\citename{Prus et~al\mbox{.} }2014]{exposmooth}
{\sc Prus, P., Borecki, M., Korwin-Pawlowski, M.~L., KociubiÅski, A., and
  Duk, M.}, 2014.
\newblock Automatic detection of characteristic points and form of optical
  signals in multiparametric capillary sensors.

\bibitem[\protect\citename{Schmidt et~al\mbox{.} }2009]{3dcurve}
{\sc Schmidt, R., Khan, A., Kurtenbach, G., and Singh, K.}
\newblock 2009.
\newblock On expert performance in 3d curve-drawing tasks.
\newblock In {\em Eurographics Symposium on Sketch-Based Interfaces and
  Modeling}, SBIM '09, 133--140.

\bibitem[\protect\citename{Stanculescu et~al\mbox{.} }2011]{freestyle}
{\sc Stanculescu, L., Chaine, R., and Cani, M.-P.}
\newblock 2011.
\newblock Freestyle: Sculpting meshes with self-adaptive topology.
\newblock {\em Comput. Graph.-UK 35}, 3, 614--622.

\bibitem[\protect\citename{Whitle }1951]{arma}
{\sc Whitle, P.}
\newblock 1951.
\newblock {\em Hypothesis Testing in Time Series Analysis}.
\newblock Statistics. Almqvist \& Wiksells.

\bibitem[\protect\citename{ZBrush }2014]{zbrush}
{\sc ZBrush}, 2014.
\newblock Zbrush.
\newblock \url{http://pixologic.com/}.

\end{thebibliography}
\end{document}